\documentclass[11pt]{article}

\usepackage{jheppub}
\usepackage{marvosym,amsmath,amssymb,bm,bbm,dsfont,cancel,tensind,stmaryrd,graphicx}
\usepackage{hyperref}
\usepackage{color}

\tensordelimiter{`}

\newcommand{\pa}{\partial}
\newcommand{\gmn}{g_{\mu\nu}}
\newcommand{\comm}[2]{\left[#1,#2\right]}
\newcommand{\set}[1]{\left\{#1\right\}}
\renewcommand{\d}{\mathrm{d}}

\newcommand{\D}{\mathrm{D}}

\newcommand{\p}[1]{\left(#1\right)}
\newcommand{\gen}[1]{\left\langle#1\right\rangle}
\newcommand{\SO}{\operatorname{SO}}
\newcommand{\ISO}{\operatorname{ISO}}
\newcommand{\nn}{\nonumber}
\newcommand{\beqn}{\begin{eqnarray}}
\newcommand{\eeqn}{\end{eqnarray}}
\newcommand{\tO}{\tilde{\omega}}
\newcommand{\te}{\tilde{e}}

\title{4D spin-2 fields from 5D Chern-Simons theory}
\author[1]{N.L. Gonz\'{a}lez Albornoz}
\author[1,2]{D. L\"{u}st}
\author[1,2]{S. Salgado}
\author[2]{A. Schmidt-May}
\affiliation[1]{Arnold Sommerfeld Center for Theoretical Physics\\
Ludwig-Maximilians-Universit\"{a}t M\"{u}nchen\\Theresienstra\ss e 37, 80333 Munich, Germany}
\affiliation[2]{Max-Planck-Institut f\"{u}r Physik\\F\"{o}hringer Ring 6, 80805 Munich, Germany}
\date{\today}

\emailAdd{nicolas.gonzalez@physik.uni-muenchen.de}
\emailAdd{dieter.luest@lmu.de}
\emailAdd{salgado@mppmu.mpg.de}
\emailAdd{angnissm@mppmu.mpg.de}

\abstract{
We consider a 5-dimensional Chern-Simons gauge theory for the isometry group of
Anti-de-Sitter spacetime, $\operatorname{AdS}_{4+1}\simeq\SO(4,2)$, and invoke different
dimensional reduction schemes in order to relate it to 4-dimensional spin-2 theories.
The AdS gauge algebra is isomorphic to a parametrized 4-dimensional conformal algebra,
and the gauge fields corresponding to the generators of non-Abelian translations 
and special conformal transformations reduce to two vierbein fields in $D=4$.
Besides these two vierbeine, our reduction schemes leave only the Lorentz 
spin connection as an additional dynamical field in the 4-dimensional theories. 
We identify the corresponding actions as particular generalizations 
of Einstein-Cartan theory, conformal gravity and ghost-free bimetric gravity in first-order form.}


\begin{document}

\begin{flushright}
LMU-ASC 75/18\\
MPP-2018-273
\end{flushright}

\maketitle
\flushbottom

\section{Introduction}\label{sec:gaugethforgrav}

General relativity (GR) is the nonlinear theory for a massless spin-2 field.
It is formulated in terms of a metric tensor whose redundant components are removed
by the diffeomorphism invariance of the Einstein-Hilbert action. This invariance is not a gauge 
symmetry \textit{\`{a} la} Lie, since the group of diffeomorphisms is not a Lie group. 
On the other hand, gauge formulations of quantum field theories seem to be a 
key ingredient for their renormalizability. For this reason, one may hope that  
a gauge formulation of gravity could help solving the problems with renormalizing GR.
As we shall briefly review below, in odd spacetime dimensions, 
$D=2n+1$, gauge formulations based on the group $\SO(2n,2)$ 
can be constructed for a certain class of consistent gravitational theories.
The corresponding actions are integrals of Chern-Simons forms which 
do not possess an analogue in even dimensions. 
Thus, in even dimensions no such construction exists for GR-like theories and one may
conclude that the problem resides in the (apparent) 4-dimensional nature of spacetime.
For pedagogical reviews on Chern-Simons formulations for gravity, we refer the
reader to Ref.~\cite{Zanelli:2005sa, Zanelli:2008sn}.

Nevertheless, there is one example for a gravitational theory in $D=4$ which 
possesses a gauge formulation: conformal gravity.
The Einstein-Hilbert action of GR is expressed in terms of the trace of the 
Riemann tensor alone; it is linear in the Ricci scalar which contains up to second derivatives 
of the metric $g_{\mu\nu}$. A famous alternative proposal is the action for the traceless 
part of the Riemann tensor; the Weyl tensor. The corresponding theory in $D=4$ which consists of only 
the square of the Weyl tensor is conformal gravity~\cite{Bach}.
Its action is invariant under an additional Weyl transformation of the metric,
$g_{\mu\nu}(x)\rightarrow \phi(x)^2g_{\mu\nu}(x)$, 
which could improve the quantum properties of gravity in principle. 
The gauge formulation for conformal gravity, based on the conformal group $\operatorname{C}_{3+1}\simeq\SO(4,2)$, 
was constructed in Ref.~\cite{kaku1977gauge}.
Conformal gravity is also a spin-2 field theory (propagating a massless and a partially
massless spin-2 mode around de Sitter backgrounds~\cite{Fradkin:1981iu, Maldacena:2011mk}) 
but it contains fourth derivatives of the 
metric and therefore leads to fatal Ostrogradski ghost instabilities~\cite{Ostrogradsky}.
For this reason, unfortunately, it cannot provide a valid theory for gravity.

The conformal gravity example suggests that extended or modified theories for gravity
may have a better chance to allow for a gauge formulation. In the ideal case this should
be possible while keeping the field content minimal, i.e.~without introducing too 
many new degrees of freedom. Maintaining the diffeomorphism symmetry 
and restricting the field content to spin-2 modes alone, 
the only known consistent extension of the Einstein-Hilbert action
is ghost-free bimetric gravity~\cite{hassan2012bimetric}. 
Its action is formulated in terms of two rank-2 tensors
with their respective Einstein-Hilbert terms and an interaction potential whose
form is strongly restricted by the absence of the Boulware-Deser ghost~\cite{Boulware:1973my}.
For a review of bimetric gravity see \cite{schmidt2016recent}. 
At the present time it is not known whether bimetric gravity possesses a gauge formulation
in $D=4$ or higher dimensions. Its relation to conformal gravity 
was discussed in Ref.~\cite{Hassan:2013pca, Hassan:2015tba, Gording:2018not}.
 
In this work we shall study the well-known Chern-Simons gauge formulation for gravity based on $\SO(4,2)$ in $D=5$. 
For earlier work on this topic see, e.g., Ref.~\cite{Edelstein:2006se, Izaurieta:2009hz}.
Here we perform several unexplored dimensional reductions in order to
unravel relations to 4-dimensional spin-2 theories, including conformal and bimetric gravity
as well as generalizations of their actions.
Kaluza-Klein reductions of Chern-Simons actions were considered in Ref.~\cite{Aros:2007nn, Morales:2017zjw} 
and alternative dimensional reductions have been proposed in 
Ref.~\cite{chamseddine1990topological, banados1997dilaton, Nastase:2007ma, Izaurieta:2012fi,
aros2014ads, Morales:2016nva}. The 5D Chern-Simons theory for the gauge group $\SO(4,2)$ shows several interesting features. The authors of Refs. \cite{Banados:1995mq,Banados:1996yj} found that its number of degrees of freedom is dependant on the location in phase space. The same happens for the generalization to $p$-form gauge connections \cite{Banados:1997qs}. In Ref. \cite{Banados:2005rz} the authors derived the holographic description and computed the Weyl anomaly. An unusual behavior on the boundary is that only the type-A anomaly is observed.

In the remainder of the introduction, before outlining our strategy and summarizing our results, 
we give a short overview of GR in first-order formulation
as well as the Chern-Simons gauge formulations for gravity in odd dimensions.


\subsection{Review of different GR formulations}

\subsubsection{Metric vs.~vielbein formulation}

The Einstein-Hilbert action of standard GR in $D=p+q$ dimensions
can be expressed in the two following ways, 
\begin{subequations}
\begin{align}
S_{\text{EH}}[g]&=\kappa\int\d^Dx\,\sqrt{-g}\,g^{\mu\nu}R_{\mu\nu}(g)\,,\label{EHact}
\\
S_{\text{EH}}[e]&=
\frac{\kappa}{(D-2)!}\int\epsilon_{a_1\ldots a_D}R^{a_1 a_2}(e)\wedge e^{a_3}\wedge\cdots\wedge e^{a_D}\,,
\label{vbEH}
\end{align}
\end{subequations}
where $\kappa$ is a constant.
The two equivalent formulations use the metric $g_{\mu\nu}$ 
or the vielbein one-form $e^a=`e^a_\mu`\d x^\mu$.
They are related via $g_{\mu\nu}(e)=`e^a_\mu``e^b_\nu`\eta_{ab}$, 
where $\eta_{ab}$ is a diagonal Minkowski metric 
with $p$ entries $+1$ and $q$ entries $-1$. 
Moreover we have used the definitions 
\begin{equation}
R(g)=g^{\mu\nu}`R^{\lambda}_{\mu\lambda\nu}`(g)\,,\qquad
R^{ab}(e)=\frac{1}{2}\,`e^a_\mu``e^b_\nu`\,`R^{\mu\nu}_{\lambda\rho}`(g(e))
\,\d x^\lambda\wedge\d x^\rho\,,
\end{equation} 
with $`R^\mu_{\nu\lambda\rho}`$ being the Riemann tensor.
Both actions $S_{\text{EH}}[g]$ and $S_{\text{EH}}[e]$ lead to the same dynamics, 
although the symmetric tensor $g_{\mu\nu}$ has $D(D+1)/2$ independent components whereas the generic vielbein
$`e^a_\mu`$ only has $D^2$. 
The metric is invariant under local Lorentz transformations (LLT), $e'^a=`\Lambda^a_b`e^b$,
where $`\Lambda^a_b`$ contains $D(D-1)/2$ independent transformation parameters. 
The additional components in $`e^a_\mu`$ can be removed by these LLT and thus
the vielbein action $S_{\text{EH}}[e]$ depends on precisely the same 
components as the metric action $S_{\text{EH}}[g]$.

\subsubsection{First-order formalism}
The first-order formalism of gravity treats connection and metric as independent objects. 
In the metric formulation one considers an independent connection $`\Gamma^\mu_{\nu\lambda}`$ 
and in the vielbein formulation one considers the spin connection 
$`\omega^{\alpha\beta}_\mu`$. The two are related via the vielbein postulate,
\begin{equation}
`\Gamma^\rho_{\mu\nu}`(\omega,e)=`e_a^\rho`\partial_\mu`e^a_\nu`+`e_a^\rho``\omega^a_{b\mu}``e^b_\nu`\,,
\end{equation}
which is the bridge between the tensor and differential form language.
In the first-order formalism the actions
\begin{subequations}
\begin{align}
S_{\text{EP}}[\Gamma,g]&=\kappa\int\d^Dx\,\sqrt{-g}\,g^{\mu\nu}R_{\mu\nu}(\Gamma)\,,\\
S_{\text{EC}}[\omega,e]&=\frac{\kappa}{(D-2)!}
\int\epsilon_{a_1\ldots a_D}R^{a_1 a_2}(\omega)\wedge e^{a_3}\wedge\cdots\wedge e^{a_D}\,,
\label{actEC}
\end{align}
\end{subequations}
are known as the Einstein-Palatini and Einstein-Cartan theory, respectively. 
The Riemann tensor now only depends on the independent connection,
\begin{align}
`R^\mu_\nu\rho\sigma`(\Gamma)&=\pa_\rho`\Gamma^\mu_\nu\sigma`-\pa_\sigma`\Gamma^\mu_\nu\rho`
+`\Gamma^\mu_\tau\rho``\Gamma^\tau_\nu\sigma`-`\Gamma^\mu_\tau\sigma``\Gamma^\tau_\nu\rho`\,,
\end{align}
and the curvature two-form is a function of the spin connection alone,
\begin{align}
R^{ab}(\omega)&=\d\omega^{ab}+`\omega^a_c`\wedge\omega^{cb}\,.
\end{align}
In the Einstein-Palatini theory, the equations of motion for $`\Gamma^\rho_{\mu\nu}`$ imply that 
the connection is the Levi-Civita connection $\Gamma(g)$, 
which is defined by the Christoffel symbols given by
\begin{align}
`\Gamma^\mu_\nu\sigma`(g)&=\tfrac{1}{2}g^{\mu\rho}\big(\partial_\nu g_{\rho\sigma}
+\partial_\sigma g_{\rho\nu}-\partial_\rho g_{\nu\sigma}\big)\,.
\end{align}
Plugging this solution back into the action, one recovers the
Einstein-Hilbert action~\eqref{EHact} in second-order form. Similarly, in the Einstein-Cartan theory, 
the equations of motion for $`\omega^{ab}_{\mu}`$ together with the vielbein postulate 
imply that $\Gamma$ is the Levi-Civita connection.

\subsection{Lanczos-Lovelock gravity} 

The $D$-dimensional Lanczos-Lovelock theory is the 
generalization of the Einstein-Hilbert (or Einstein-Cartan) action 
to the most general polynomial in $R^{ab}$ and $e^a$ built with 
the invariant tensor for the Lorentz algebra coming from the Euler class \cite{lanczos1932elektromagnetismus,lanczos1938remarkable,lovelock1971einstein,troncoso2000higher}. 
For instance, in $D=4$ the Lanczos-Lovelock action reads
\begin{equation}
S_{\text{LL},4}[\omega,e]=\int\epsilon_{abcd}\Big(a_{20}\,R^{ab}\wedge R^{cd}
+a_{12}\,R^{ab}\wedge e^c\wedge e^d+a_{04}\,e^a\wedge e^b\wedge e^c\wedge e^d\Big)\,,
\label{eq:4dll}
\end{equation}
where $a_{mn}$ are constants multiplying the term of $m$ order in $R^{ab}$ and $n$ order in $e^a$. 
The term proportional to $a_{04}$ is the cosmological constant term in 4 dimensions.
Accordingly, in general dimension $D$, the cosmological term is the polynomial of order $D$ in the vielbein. 
The action contains a term quadratic in the curvature, known as the Gauss-Bonnet term,
which is topological in $D=4$. Its generalization to any even dimension $D=2n$ is the 
topological $a_{n0}$ term consisting only of curvatures. 

The gravitational actions we considered so far are generically invariant under 
LLT but not under the entire set of Poincar\'{e} transformations,
since there is no translational symmetry generator which would correspond to the vielbein.
An exception is a particular type of Lanczos-Lovelock gravity 
in odd dimensions $D=2n+1$, which can be constructed 
as a gauge theory for the ($2n+1$)-dimensional AdS group 
$\operatorname{AdS}_{2n+1}\simeq\SO(2n, 2)$. 
The gauge connection $\bm{A}$ of $\SO(2n,2)$ can be decomposed as follows,
\begin{equation}
\bm{A}=\frac{1}{2}\omega^{AB}\bm{J}_{AB}=\frac{1}{2}\omega^{ab}\bm{J}_{ab}+e^a\bm{P}_{a}\,.
\qquad A,B=1,\ldots, D+1\,,
\end{equation}
where one identifies the component $\omega^{a,D+1}$ with the vielbein $e^a$.
The generators satisfy the algebra
\begin{align}\label{eq:adsalgebra1} 
\comm{\bm{J}_{ab}}{\bm{J}_{cd}}=`f_{ab,cd}^{ef}`\bm{J}_{ef}\,,\qquad
 \comm{\bm{J}_{ab}}{\bm{P}_{c}}=`f_{ab,c}^{d}`\bm{P}_{d}\,,\qquad
\comm{\bm{P}_{a}}{\bm{P}_{b}}=-\ell^{-2}\bm{J}_{ab}\,,
\end{align}
where $\ell$ is a constant with dimensions of length. 
Using this gauge algebra to construct a $(2n+1)$-dimensional Chern-Simons theory 
and decomposing the indices $A=(a,D+1)$ as above,
one recovers the $D$-dimensional Lanczos-Lovelock Lagrangian 
with some particular coefficients $a_{mn}$ \cite{troncoso2000higher}. 
The corresponding action was first obtained in $D=3$~\cite{witten19882+} 
and later in all odd dimensions~\cite{Chamseddine:1989nu}.

In the commutation relations of the $\SO(4,2)$ algebra the translations $\bm{P}_a$ should 
not be understood in a Poincar\'{e} sense 
since here $\comm{\bm{P}_a}{\bm{P}_b}\neq0$. 
However, the Poincar\'{e} symmetry $\operatorname{P}_{2n+1}$ is recovered at the level of the 
algebra after performing an In\"{o}n\"{u}-Wigner (IW) contraction $\ell\rightarrow\infty$ 
as we can see from \eqref{eq:adsalgebra1}. 
It follows that a Chern-Simons gauge theory for the Poincar\'{e} group
in $D=2n+1$ is given by~\cite{banados1996higher}
\begin{align}
S=a_{n1}\int\epsilon_{a_1\ldots a_{2n+1}}\,R^{a_1a_2}(\omega)\wedge\cdots\wedge R^{a_{2n-1}a_{2n}}(\omega)\wedge e^{a_{2n+1}}\,,
\label{eq:5dcspoincare}
\end{align}
which for $D>3$ clearly differs from the Einstein-Cartan action~\eqref{actEC} 
which only contains one power of curvature.

In even dimensions there still exists no gauge formulation 
for the Einstein-Cartan or the Lanczos-Lovelock theory. 
The question thus remains whether it is possible to derive standard GR in even dimensions 
from a gauge theory for the Poincar\'{e} or for a more general group. 
Such an attempt was made in Ref.~\cite{banados1997dilaton} which
showed that $D=4$ dilaton Einstein gravity is obtainable from a 
$D=5$ Chern-Simons theory under certain assumptions.

\subsection{Outline of approach and summary of results} 

Inspired by the results for the Lanczos-Lovelock theory, we will address
the question whether it is possible to derive a 4-dimensional spin-2 field theory 
from a pure gauge formulation in $D=5$. Clearly, this setup will involve a dimensional
reduction of the field theory.

More precisely, we will explore relations among the Chern-Simons gauge theory 
for the AdS group $\operatorname{AdS}_{4+1}$ in $D=5$ and spin-2 field theories in $D=4$ 
which have the following origin.
The group $\operatorname{AdS}_{4+1}\simeq \SO(4,2)$ generated by rotations 
$\bm{J}_{AB}$ and translations $\bm{T}_A$ with $A,B=1,\hdots,5$, is isomorphic to the 4-dimensional 
conformal group $\operatorname{C}_{3+1}$ generated by 
rotations $\bm{J}_{ab}$, translations $\bm{P}_a$, conformal transformations 
$\bm{K}_a$ and dilatations $\bm{D}$ with $a,b=1,\hdots,4$. 
After dimensional reduction, the generators $\bm{J}_{ab}$ will give rise to the 
spin connection of the 4-dimensional LLT. The generators $\bm{P}_a$ and $\bm{K}_a$,
on the other hand will introduce two 4-dimensional vierbein fields.
Thus, in the general case we will not arrive at GR in $D=4$ 
but recover theories involving two spin-2 fields.

We study the algebra of the gauge group $\SO(4,2)$ in different bases
and thereby identify the isomorphism to the algebra of $\operatorname{C}_{3+1}$. 
Even though the 5-dimensional Chern-Simons actions expressed in
different bases of the algebra are all related to each other by linear field redefinitions,
our dimensional reduction scheme is basis dependent and therefore 
leads to inequivalent theories in $D=4$.
Generically, the reduction breaks the gauge symmetry down to 
$\SO(3,1)$, corresponding to the 4-dimensional LLT. 
In this way, we recover the following 4-dimensional spin-2 theories.
\begin{itemize}

\item {\bf Einstein-Cartan theory:} We perform a simple dimensional reduction after 
taking an IW contraction limit that reduces $\SO(4,2)$ to $\ISO(3,2)$ or $\ISO(4,1)$. 
This results in Einstein-Cartan theory in $D=4$ plus a Lorentz-breaking term involving torsion
which can be removed by restricting a field in the 5-dimensional action.

\item {\bf First-order conformal gravity:} Without taking the IW contraction limit, 
a very similar dimensional reduction of the Chern-Simons action results in a 
first-order formulation of conformal gravity in $D=4$. 
An identical action was first obtained in a different setup in Ref.~\cite{kaku1977gauge}. 
The Weyl symmetry of our 4-dimensional action is not part of the original gauge group
which is broken to $\SO(3,1)$ by the dimensional reduction. Instead it originates from
an $\SO(1,1)$ symmetry of the gauge algebra, which acts on the vierbein fields
as a mixture of a rotation and a Weyl transformation.

\item {\bf Generalized first-order conformal gravity:} We consider 
a different dimensional reduction scheme in a particular basis and 
introduce two warp functions, one for each vielbein. 
Integrating along the warp direction reduces the Chern-Simons action to an effective theory 
with more free parameters than in the previous case. 
Our result can thus be viewed as a generalization of first-order conformal gravity. 

\item {\bf Generalized first-order bimetric theory:} We consider two copies of the 
Chern-Simons action in $D=5$ with interdependent field content which explicitly breaks
 the gauge symmetry down to $\SO(3,1)\times\SO(2)$. 
 Dimensionally reducing this action 
 results in a generalized bimetric theory involving a new type of kinetic interaction. 
These novel derivative terms can be removed  by restricting fields in the 5-dimensional action,
in which case we obtain the standard bimetric theory \emph{\`{a} la} Hassan and Rosen.
This further breaks the gauge group down to $\SO(3,1)$. 
We also discuss another type of restriction which recovers the Weyl rotation symmetry of
first-order conformal gravity.

\end{itemize}

\paragraph{Organization of this manuscript.}
Section~\ref{sec:cgb} briefly reviews 4-dimensional 
conformal gravity and bimetric gravity in the language of differential forms.
In section~\ref{sec:confgroup} we derive the commutation relations of the 
conformal algebra $\operatorname{C}_{3+1}$ 
and its parameterized version $\operatorname{C}_{3+1}(M,\gamma)$
starting from the special orthogonal algebra $\SO(4,2)$.
Section~\ref{sec:csgeometry} formulates a 5-dimensional Chern-Simons gauge theory 
for the group $\SO(4,2)\simeq\operatorname{C}_{3+1}(M,\gamma)$.
In Sec. \ref{sec:csdimred} we analyze different dimensional reductions 
of the Chern-Simons action expressed in different bases of the gauge algebra. 
We discuss our findings in section~\ref{sec:discussion}. The appendix contains
some technical details of the gauge theory and explicit expressions
of the Chern-Simons action.

\section{Conformal and bimetric gravity in first-order form}\label{sec:cgb}

Since our constructions will produce (generalized) versions of conformal and bimetric gravity 
formulated in terms of vierbein fields, we briefly review these setups in the following.

\subsection{First-order version of conformal gravity}\label{sec:cgc}

Let us start from the following action for two vielbein 1-forms $e^a$ and $h^a$ in $D=4$, 
\begin{align}\label{eq:focg}
\nn S[e,h]&=\int_{M_4}\epsilon_{abcd}\Big(R^{ab}(\omega)
\wedge\Big[e^{c}\wedge e^{d}-\alpha^2h^{c}\wedge h^{d}\Big]\\
&\qquad+\tfrac{\eta}{2}\Big[e^{a}\wedge e^{b}-\alpha^2h^{a}\wedge h^{b}\Big]
\wedge\Big[e^{c}\wedge e^{d}-\alpha^2h^{c}\wedge h^{d}\Big]\Big)\,,
\end{align} 
where $`\omega^{ab}_\mu`=`\omega^{ab}_\mu`(e+\alpha h)$ is the spin connection
of a linear combination of the two vierbeine. It is related to the Levi-Civita connection 
$\Gamma(g)$ of the metric $g=(e+\alpha h)^\mathrm{T}\eta(e+\alpha h)$ via the vielbein postulate. 
In order to see the relation of this action to conformal gravity, 
we define the following linear combinations,
\begin{equation}
E^a=e^a+\alpha h^a\,,\qquad H^a=e^a-\alpha h^a\,,
\end{equation}
as well as the tensors
\beqn\label{deftensors}
g_{\mu\nu}\equiv \eta_{ab}E^a_{~\mu}E^b_{~\nu}\,,\qquad
S_{\mu\nu}\equiv \eta_{ab}E^a_{~\mu}H^b_{~\nu}\,,\qquad
S^{\mu\nu}\equiv g^{\mu\rho}g^{\nu\sigma}S_{\rho\sigma}\,.
\eeqn
These satisfy,
\beqn
H^a=H`{}^a_\mu`\d x^\mu=`E^a_{\mu}`S`{}^\mu_\nu`\d x^\nu.
\eeqn
Using the vierbein postulate, we can express the curvature two-form as,
\beqn
R^{ab}(e+\alpha h)=\frac{1}{2}`R^{ab}_{\rho\sigma}`(E)\,\d x^\rho\wedge \d x^\sigma
=\frac{1}{2}`E^a_{\mu}``E^b_{\nu}``R^{\mu\nu}_{\rho\sigma}`(g)\,\d x^\rho\wedge \d x^\sigma\,.
\eeqn
In terms of the tensors, the action can then be brought into the following form,
\begin{align}\label{cgaux2}
S[g,S]=\int\d^4x\sqrt{-g}\,&\Big(`S^\mu_\mu`R(g)-2S^{\mu\nu}R_{\mu\nu}(g)
 -2m^2\Big[(`S^\mu_\mu`)^2-S^{\mu\nu}S_{\mu\nu}\Big]\Big)\,.
\end{align}
Upon integrating out $S_{\mu\nu}$ we finally obtain the action for conformal gravity,
\begin{align}\label{cgbim}
S[g]=\int\d^4x~\sqrt{-g}\Big( R^{\mu\nu}(g)R_{\mu\nu}(g)-\tfrac{1}{3}R(g)^2\Big)\,,
\end{align}
which is invariant under Weyl scalings of the metric, $\gmn\mapsto \phi(x)^2\gmn$. 
This discussion shows that conformal gravity can be written in the equivalent form (\ref{eq:focg}).

Note that from the above considerations,  
it is not obvious how to rewrite conformal gravity in an equivalent first-order formulation.  
The reason is that the spin connection $`\omega^{ab}_\mu`$ defining the curvature is that 
of the vierbein $E^a_{~\mu}$ which is not a solution to its equations
of motion when $`\omega^{ab}_\mu`$ is treated as an independent field in $R^{ab}(\omega)$.

The action (\ref{eq:focg}) with independent spin connection was previously
obtained in Ref.~\cite{kaku1977gauge} from a 4-dimensional gauge theory for the conformal group.
It turned out to be invariant under proper conformal transformations only if the spin
connection is subject to a constraint which sets $`\omega^{ab}_\mu`$ equal to
$`\omega^{ab}_\mu`(e+\alpha h)$ plus a correction term, which however drops out from the final
action when the auxiliary tensor field has been integrated out. The resulting action is precisely
conformal gravity as in (\ref{cgbim}).

The action (\ref{eq:focg}) with arbitrary $`\omega^{ab}_\mu`$
is invariant under an $\SO(1,1)$ rotation of the vierbein vector,
\begin{equation}\label{weylrot}
\left(\begin{array}{c}
E'^a(x)\\
H'^a(x)
\end{array}\right)=\left(\begin{array}{cc}
                          \cosh\phi(x) &\ \sinh\phi(x)\\
                          \sinh\phi(x) &\ \cosh\phi(x)
                         \end{array}
\right)\left(\begin{array}{c}
E^a(x)\\
H^a(x)
\end{array}\right)\,,
\end{equation}
which we will refer to as a \emph{Weyl rotation} from now. 
The infinitesimal version of a $\SO(1,1)$ Weyl rotation is given by
\begin{equation}
\label{eq:wrd} \delta E^a(x)=\phi(x)\,H^a(x)\,,\qquad \delta H^a(x)=\phi(x)\,E^a(x)\,.
\end{equation}
From the infinitesimal transformations \eqref{eq:gtdcb} with $\theta^{ab}=\rho^a=b^a=0$,
we see that the Weyl rotation \eqref{eq:wrd} is also a subgroup of $\SO(4,2)$.

\subsection{First-order formulation of bimetric gravity}\label{sec:bgc}

Bimetric gravity consists of two Einstein-Hilbert terms and an 
interaction term for two independent metrics. 
The interactions need to be chosen in a particular way such that they avoid
classical ghost instabilities \cite{hassan2012bimetric}. 
The vielbein formulation of bimetric gravity was introduced in Ref.~\cite{hinterbichler2012interacting}. 
Going to the first-order formalism is straightforward: one simply introduces one spin connection 
for each metric using two copies of the vielbein postulate. 
In $D=4$, bimetric gravity in the Cartan formalism then reads 
\begin{align}\label{BGact}
\nn S_{\text{BGC}}[\omega,\tilde\omega,e,\tilde e]
\nn &=\tfrac{1}{2}\int\epsilon_{abcd}\Big(m_e^{2}\,R^{ab}(\omega)\wedge e^{c}\wedge e^{d}+m_{\tilde{e}}^{2}\,R^{ab}(\tilde\omega)\wedge\tilde{e}^{c}\wedge\tilde{e}^{d}\Big)\\
&\quad-\tfrac{m^4}{8}\int\epsilon_{a_1a_2a_3a_4}\textstyle\sum_{n=0}^4\frac{\beta_n}{n!(4-n)!}\,e^{a_1}\wedge\cdots\wedge e^{a_n}\wedge \tilde{e}^{a_{n+1}}\wedge\cdots\wedge\tilde{e}^{a_4}\,.
\end{align}
The equations of motion for $\omega^{ab}$ and $\tilde\omega^{ab}$ together with 
two vielbein postulates for the connections $\Gamma$ and $\tilde\Gamma$
imply that they are the Christoffel connections for the metrics 
associated to $`e^a_\mu`$ and $\tilde{e}`{}^a_\mu`$. Thus the kinetic terms become Einstein-Hilbert.
The interaction terms can also be expressed in terms of the corresponding metric tensors
by making use of the constraint $\eta_{ab}`e^a_{[\mu}`\tilde{e}`{}^b_{\nu]}`=0$,
which follows from the equations of motion.

A special bimetric model was studied in Ref.~\cite{hassan2013partially}. 
It corresponds to a certain choice for
the $\beta_n$ parameters in (\ref{BGact}) for which the action assumes the form
\begin{align}\label{genact}
S_{\operatorname{PM}}[\omega,\tO,e,\te]
&=m_e^2\int \epsilon_{abcd}\Big( R^{ab}(\omega)\wedge e^c\wedge e^d
+\alpha^2 R^{ab}(\tO)\wedge \te^c\wedge \te^d\nn\\
&\qquad\qquad\qquad-m^2\Big[e^a\wedge e^b+\alpha^2\te^a\wedge \te^b\Big]
\wedge\Big[e^c\wedge e^d+\alpha^2\te^c\wedge \te^d\Big]\Big)\,.
\end{align}
Interestingly, this model shows some similarities to conformal gravity.
Namely, if we formally set the two spin connections to be equal, 
$\tilde{\omega}`{}^{ab}_{\mu}`=\omega`{}^{ab}_{\mu}`$, and
take $\alpha\rightarrow i\alpha$, we recover the action~\eqref{eq:focg} for conformal gravity in the vielbein 
formalism.\footnote{Note that the spin-2 ghost of conformal gravity is a consequence of making 
$\alpha$ imaginary which introduces a minus sign in front of a kinetic term.} 
Moreover, the action \eqref{genact} has been shown to reproduce conformal gravity to lowest order
in a curvature expansion~\cite{Hassan:2013pca, Hassan:2015tba, Gording:2018not}

An interesting approach was taken by the authors of Ref.~\cite{apolo2017non}
who started from a 4-dimensional gauge theory for the group $\SO(4,1)$ to construct
an action for two spin-2 fields interacting with a vector field. The resulting action
contains the same potential as in \eqref{genact} but, due to a modified kinetic structure 
and the interactions with the vector field, the theory possesses a residual
$\SO(3,1)\times \SO(2)$ invariance. Whether it is free from ghost instabilities is an open question.

\section{Parametrized conformal algebra}\label{sec:confgroup}
It is well-known that the special orthogonal group $\SO(4,2)$
used in the construction of a gauge theory for gravity in $D=5$
is isomorphic to the conformal group $\operatorname{C}_{3+1}$ in $D=4$~\cite{ammon2015gauge}. 
In the following we start from the algebra of $\SO(4,2)$,
for which we introduce new sets of bases with the aim to exhibit the isomorphism to the conformal algebra.

\subsection{Bases of the algebra}\label{sec:confalg}

\subsubsection{6-covariant basis}
We start from the basis in which the algebra 
of the antisymmetric generators $\set{\bm{J}_{IJ}}$ of $\SO(4,2)$ 
are expressed in a covariant manner. The commutation relations are  
\begin{equation}\label{so42algebra}
\comm{\bm{J}_{IJ}}{\bm{J}_{KL}}=`f_{IJ,KL}^MN`\bm{J}_{MN}\,,\qquad I,J,\ldots=1,\ldots,6\,,
\end{equation}
with structure constants
\begin{equation}
`f_{IJ,KL}^MN`=-\frac{1}{2}\Big(\eta_{IK}\delta_{JL}^{MN}+\eta_{JL}\delta_{IK}^{MN}-\eta_{JK}\delta_{IL}^{MN}-\eta_{IL}\delta_{JK}^{MN}\Big)
\end{equation}
and $\eta_{IJ}=\operatorname{diag}(+,+,+,-,-\eta,\eta)$ with $\eta=\pm1$. 
We call $\set{\bm{J}_{IJ}}$ the \textit{6-covariant basis} and $I,J,\ldots$ \textit{6-covariant internal indices}. 
In this basis the invariant tensor of the Euler class is given by
\begin{equation}
\label{eq:invten6covbasis} \gen{\bm{J}_{IJ},\bm{J}_{KL},\bm{J}_{MN}}=\epsilon_{IJKLMN}\,.
\end{equation}
With this we have all the ingredients we need to compute the gauge CS theory for the $\SO(4,2)$.

\subsubsection{5-covariant basis}
Expanding indices we can express the commutation relations in a 5-covariant basis. 
To this end, we take an anti-symmetric sub-matrix of $\bm{J}_{IJ}$ and a scaled vector, 
namely $\bm{J}_{AB}$, $\bm{J}_{A6}\equiv\gamma\,\bm{T}_A$ with $A,B,\ldots=1,\ldots,5$.
The real constant $\gamma$ has been introduced to perform an IW contraction later on. 
The commutation relations (\ref{so42algebra}) of 
$\SO(4,2)$ in terms of these 5-covariant generators read
\begin{subequations}\label{eq:commrelso42-5cov} 
\begin{align}
\label{eq:commrelso42-5cov1} \comm{\bm J_{AB}}{\bm J_{CD}}&=`f_{AB,CD}^{EF}`\bm J_{EF}\,,\\
\label{eq:commrelso42-5cov2} \comm{\bm J_{AB}}{\bm T_{C}}&=`f_{AB,C}^D`\bm T_D\,,\\
\label{eq:commrelso42-5cov3} \comm{\bm T_{A}}{\bm T_{B}}&=\gamma^{-2}`f_{A,B}^CD`\bm J_{CD}\,,
\end{align}
\end{subequations}
with structure constants
\begin{align}
`f_{AB,C}^D`=-\Big(\eta_{AC}\delta^D_B-\eta_{BC}\delta^D_A\Big),\qquad
`f_{A,B}^CD`=-\tfrac{1}{2}\,\eta\,\delta_{AB}^{CD},
\end{align}
and where $\eta_{AB}=\operatorname{diag}(+,+,+,-,-\eta)$ with $\eta=\pm 1$.
We call $A,B,\ldots$ 
\textit{5-covariant internal indices}. In this basis the only non-zero 
invariant tensor of the Euler class is given by
\begin{equation}
\gen{\bm{J}_{AB},\bm{J}_{CD},\bm{T}_E}=\gamma^{-1}\,\epsilon_{ABCDE}\,.
\label{eq:invten5basis}
\end{equation}
The commutation relations~\eqref{eq:commrelso42-5cov} show that the IW 
contraction $\gamma\rightarrow\infty$ converts the algebra 
into one of the 5-dimensional Poincar\'{e} algebras, $\ISO(3,2)$ or 
$\ISO(4,1)$ for $\eta=\pm1$, respectively.

\subsubsection{4-covariant basis}
We continue with expanding indices. Consider now the generators 
$\bm{J}_{ab}$, $\bm{J}_{a5}\equiv\bm{B}_a$, $\bm{T}_a$, $\bm{T}_{5}\equiv\bm{D}$
which make up the \textit{4-covariant basis} whose 4-covariant indices
 $a,b,\ldots$ will be interpreted as local Lorentz indices. 
The commutation relations (\ref{so42algebra}) in this basis read
\begin{subequations}
\begin{align}
\comm{\bm{J}_{ab}}{\bm{J}_{cd}}&=`f_{ab,cd}^{ef}`\,\bm{J}_{ef}\,,\\
\comm{\bm{J}_{ab}}{\bm{B}_c}&=`f_{ab,c}^d`\,\bm{B}_d\,,\\
\comm{\bm J_{ab}}{\bm{T}_c}&=`f_{ab,c}^{d}`\bm{T}_d\,,\\
\comm{\bm J_{ab}}{\bm{D}}&=0\,,\\
\comm{\bm{B}_{a}}{\bm{B}_{b}}&=\eta\,\bm{J}_{ab}\,,\\
\comm{\bm{B}_{a}}{\bm{T}_b}&=-\eta_{ab}\,\bm{D}\,,\\
\comm{\bm{B}_{a}}{\bm{D}}&=-\eta\,\bm{T}_a\,,\\
\comm{\bm{T}_{a}}{\bm{T}_{b}}&=-\eta\gamma^{-2}\bm{J}_{ab}\,,\\
\comm{\bm{T}_{a}}{\bm{D}}&=-\eta\gamma^{-2}\,\bm{B}_{a}\,.
\end{align}
\end{subequations}
The corresponding gauge transformations are listed in appendix~\ref{app:gt4c}.
We note that the above algebra is invariant under $\SO(1,1)$ rotations\footnote{For
$\gamma^2<0$, the symmetry would be $\SO(2)$.} of the vector
$(\bm{B}^a, \gamma \bm{T}^a)$,
\beqn\label{algrot}
\begin{pmatrix}
\bm{B}^a\\
\gamma \bm{T}^a
\end{pmatrix}
\longmapsto
\begin{pmatrix}
\cosh\phi&\sinh\phi\\
\sinh\phi&\cosh\phi
\end{pmatrix}
\begin{pmatrix}
\bm{B}^a\\
\gamma \bm{T}^a
\end{pmatrix}\,,
\eeqn
which is reminiscient of the Weyl rotation in equation (\ref{weylrot}). 
Indeed, we will identify the symmetry of the algebra as the origin of the Weyl rotation invariance
later on.

In the above 4-covariant basis the isomorphism between the 
$\SO(4,2)$ and the 4-dimensional conformal algebra is still not explicit. 
Only the subalgebra $\operatorname{span}\set{\bm{J}_{ab},\bm{D}}$ 
is the same as in the conformal algebra. To retrieve the full isomorphism to the conformal algebra 
we will perform a general linear transformation on the generators $\bm{B}_a$ and 
$\bm{T}_a$ in the following section.

\subsubsection{4-covariant canonical basis and parametrized conformal algebra}
\label{sec:4covcanbas}
We now introduce a new basis for the subspace spanned by the generators $\bm{B}_a$ and $\bm{T}_a$, 
and define the new generators $\bm{P}_a$ and $\bm{K}_a$ as the following
$\operatorname{GL}(2,\mathbbm{C})$ rotated combinations, 
\begin{equation}
\left(\begin{array}{c}
\bm{P}_a\\
\bm{K}_a
\end{array}\right)=M\left(\begin{array}{c}
\bm{B}_a\\
\gamma \bm{T}_a
\end{array}\right)\,,\qquad M=\left(\begin{array}{cc}
a & \ b\gamma^{-1}\\
 c & \  d\gamma^{-1}
\end{array}\right)\,, 
\qquad a,b,c,d\in \mathbb{C}\,.
\label{eq:rotation12}
\end{equation}
The full commutation relations in this new basis read
\begin{subequations}
\begin{align}
\comm{\bm J_{ab}}{\bm J_{cd}}&=`f_{ab,cd}^{ef}`\,\bm J_{ef}
\label{alg1} \,,\\
\comm{\bm J_{ab}}{\bm P_{c}}&=`f_{ab,c}^{d}`\,\bm P_{d}
\label{alg2}\,,\\
\comm{\bm J_{ab}}{\bm K_{c}}&=`f_{ab,c}^{d}`\,\bm K_{d}
\label{alg3}\,,\\
\comm{\bm J_{ab}}{\bm{D}}&=0
\label{alg4}\,,\\
\label{eq:confabcdpp} \comm{\bm P_{a}}{\bm P_{b}}&=\eta\,\Big(a^2-b^2\gamma^{-2}\Big)\bm J_{ab}\,,\\
\comm{\bm P_{a}}{\bm K_b}&=-\det M\,\eta_{ab}\,\bm{D}+\eta\,\Big(ac-bd\gamma^{-2}\Big)\bm J_{ab}\,,\\
\label{eq:confabcdpd} \comm{\bm P_{a}}{\bm{D}}&=\tfrac{\eta}{\det M}\Big[\Big(ac-bd\gamma^{-2}\Big)\bm{P}_a-\big(a^2-b^2\gamma^{-2}\Big)\bm{K}_a\Big]\,,\\
\label{eq:confabcdkk} \comm{\bm K_a}{\bm K_b}&=\eta\,\Big(c^2-d^2\gamma^{-2}\Big)\bm J_{ab}\,,\\
\label{eq:confabcdkd} \comm{\bm K_a}{\bm{D}}&=\tfrac{\eta}{\det M}\Big[\Big(c^2-d^2\gamma^{-2}\Big)\bm{P}_a-\Big(ac-bd\gamma^{-2}\Big)\bm{K}_a\Big]\,.
\end{align}
\end{subequations}
Since it represents the most general form of linearly redefined translational and conformal sectors,
we call this the \textit{4-covariant canonical basis}.
The corresponding gauge transformations are listed in appendix~\ref{app:gt4cc}.
The algebra remains invariant under the $\SO(1,1)$ rotations of the vector $(\bm{B}^a, \gamma \bm{T}^a)$,
which by inverting the matrix $M$ can be expressed in terms of $\bm{P}^a$ and~$\bm{K}^a$.
 
As we will see below, the above set of isomorphic algebras with general $M$
contains the conformal algebra as a special case,
and we therefore call it the \textit{parametrized conformal algebra} $\operatorname{C}_{3+1}(M,\gamma)$. 
It contains 8 independent real parameters in total.

\subsection{Interesting parameter choices}\label{sec:choosingcoefc}

We now identify two special choices for the parameters $a$, $b$, $c$, $d$ and 
$\gamma$, which will lead to different but isomorphic subalgebras. 
The first four commutators (\ref{alg1})-(\ref{alg4}) of the algebra are independent of these
parameters and thus remain unaffected.
We can immediately rule out parameter choices which render the matrix $M$ in eq.~(\ref{eq:rotation12}) 
singular, since for these the commutators 
\eqref{eq:confabcdpd} and \eqref{eq:confabcdkd} remain undetermined.
The invertibility of the matrix $M$ also ensures that the change of basis in eq.~(\ref{eq:rotation12}) 
is invertible and therefore all the algebras $\operatorname{C}_{3+1}(M,\gamma)$ are isomorphic.

\subsubsection{Conformal basis}
In this paper we will always treat the generators $\bm{P}_a$ and $\bm{K}_a$ on the same footing 
since we will interpret the corresponding gauge fields as two vierbeine with equal status.
We therefore start by analyzing the case where both subspaces $\operatorname{span}\set{\bm{P}_a}$ and 
$\operatorname{span}\set{\bm{K}_a}$ each make up an Abelian subalgebra. 
One could also analyze cases with only one Abelian subalgebra which we will
not cover in the following. The double Abelian case is obtained when
\begin{equation}
a^2-b^2\gamma^{-2}=0\,,\qquad c^2-d^2\gamma^{-2}=0\,.
\label{eq:case1}
\end{equation}
Note that this implies that the matrix $M$ maps onto a basis of two vectors 
$(a,b\gamma^{-1})$ and $(c, d\gamma^{-1})$ which have zero-norm with respect to the $\SO(1,1)$ 
invariant metric.
The above equations can be satisfied by two distinct parameter choices: 
{\bf i)} $b=\pm a\gamma$, $d=\pm c\gamma$ or {\bf ii)}
$b=\pm a\gamma$, $d=\mp c\gamma$.
The case {\bf i)}  implies $\det M=0$, so we rule it out. 
With {\bf ii)}  on the other hand we get $\det M=\mp2ac\gamma\neq 0$.  
As mentioned above, the first four commutators of the algebra are identical to
(\ref{alg1})-(\ref{alg4}). The remaining ones for the allowed choice {\bf ii)} read
\begin{subequations}\label{eq:confalg} 
\begin{align}
\comm{\bm P_{a}}{\bm P_{b}}&=\comm{\bm K_{a}}{\bm K_{b}}=0\,,\\
\label{eq:ccpk} \comm{\bm P_{a}}{\bm K_{b}}&=\pm2ac\p{\gamma\eta_{ab}\,\bm{D}\pm\eta\bm J_{ab}}\,,\\
\comm{\bm P_{a}}{\bm{D}}&=\mp\eta\gamma^{-1}\bm{P}_{a}\,,\\
\label{eq:confalg9} \comm{\bm K_{a}}{\bm{D}}&=\pm\eta\gamma^{-1}\bm{K}_{a}\,.
\end{align}
\end{subequations}
We call this choice of parameters the \textit{4-conformal basis}. 
In this case the algebra still depends on 3 real parameter combinations
and it contains the 4-dimensional Poincar\'{e} subalgebra
\begin{equation}
\operatorname{P_{3+1}}\simeq\operatorname{span}\set{\bm{J}_{ab},\bm{P}_a}\simeq\operatorname{span}\set{\bm{J}_{ab},\bm{K}_a}\,.
\end{equation}
By re-absorbing $\gamma$ into $\bm{D}$ and 
furthermore choosing $ac=1$, we obtain the algebra of the conformal group $\operatorname{C}_{3+1}$,
written in the usual basis of conformal field theory.

\subsubsection{Orthogonal basis}

Another special case which simplifies the algebra is obtained by demanding that $M$
maps onto a set of vectors $(a,b\gamma^{-1})$ and $(c, d\gamma^{-1})$ which are orthogonal 
with respect to the $\SO(1,1)$ invariant metric. This requires
\begin{equation}
ac-bd\gamma^{-2}=0\,.
\label{eq:case2}
\end{equation}
This leaves $M$ non-singular.
In this case the first four commutators of the algebra are again identical to
(\ref{alg1})-(\ref{alg4}).
The remaining part of the algebra reads
\begin{subequations}
\begin{align}
\label{eq:adsconfbasispp} \comm{\bm P_{a}}{\bm P_{b}}&=\eta\p{a^2-b^2\gamma^{-2}}\bm J_{ab}\,,\\
\comm{\bm P_{a}}{\bm K_{b}}&=-\det M\,\eta_{ab}\,\bm{D}\,,\\
\label{eq:adsconfbasispd} \comm{\bm P_{a}}{\bm{D}}&=-\tfrac{\eta}{\det M}\p{a^2-b^2\gamma^{-2}}\bm{K}_{a}\,,\\
\label{eq:adsconfbasiskk} \comm{\bm K_{a}}{\bm K_{b}}&=\eta\p{c^2-d^2\gamma^{-2}}\bm J_{ab}\,,\\
\label{eq:adsconfbasiskd} \comm{\bm K_{a}}{\bm{D}}&=\tfrac{\eta}{\det M}\p{c^2-d^2\gamma^{-2}}\bm{P}_{a}\,,
\end{align}
\end{subequations}
with $bd= ac\gamma^2$. The algebra still depends on 4 real parameter combinations and
we call it the \textit{orthogonal basis}. The corresponding gauge transformations are listed in appendix~\ref{app:gt4dc}.

\section{Chern-Simons geometry in $D=5$}\label{sec:csgeometry}

We now set out to formulate a gauge theory for the 4-dimensional conformal group 
based on a 5-dimensional Chern-Simons action.

\subsection{Chern-Simons action in 6-covariant basis}

Let us consider the gauge theory defined by a Chern-Simons form $Q^{(5)}$ on a 5-dimensional manifold,
\begin{equation}\label{eq:csaction}
S_{\operatorname{CS}_5}[\bm{A}]=\int_{M_5}Q^{(5)}(\bm{A})
=\int_{M_5}\Big\langle\bm{F}\wedge\bm{F}\wedge\bm{A}-\tfrac{1}{2}\bm{F}\wedge\bm{A}\wedge\bm{A}\wedge\bm{A}+\tfrac{1}{10}\bm{A}\wedge\bm{A}\wedge\bm{A}\wedge\bm{A}\wedge\bm{A}\Big\rangle\,,
\end{equation}
where the gauge connection  $\bm{A}=\tfrac{1}{2}{A}^{IJ}\bm{J}_{IJ}$  
is valued in the algebra of the conformal group 
$\operatorname{C}_{3+1}\simeq\SO(4,2)$.  
The action is invariant under the conformal group by construction.  
It defines a 5-dimensional geometry specified by a connection instead of a metric. 
In the following, lower-dimensional metrics will emerge through the subspace decomposition 
of the gauge connection.

\subsection{5-covariant basis}
In the 5-covariant basis the gauge connection
splits into a connection $\omega^{AB}={ A}^{AB}$ with $A,B=1,\hdots 5$ of
the rotational group $\SO(3,2)$ or $\SO(4,1)$ 
and an additional gauge field $u^A={ A}^{A6}/\gamma$ 
associated to the generator $\bm{T}_A=\gamma\bm{J}_{A6}$,
\begin{equation}\label{connectionsplit}
\bm{A}=\tfrac{1}{2}{A}^{IJ}\bm{J}_{IJ}
=\bm{\omega}_5+\bm{u}_5=\tfrac{1}{2}\omega^{AB}\bm{J}_{AB}+u^{A}\bm{T}_{A}\,.
\end{equation}
The form of the action (\ref{eq:csaction}) in terms of 5-covariant objects can be computed using
the subspace separation method of Ref.~\cite{izaurieta2007extended}.
The result is
\begin{align}\label{eq:csforso42}
\nn S_{\text{CS}_5}[\omega_5,u_5]
&=\tfrac{1}{4}\int_{M_5}\epsilon_{ABCDE}\Big(R^{AB}_5(\omega_5)\wedge R^{CD}_5(\omega_5)\wedge u^{E}-\tfrac{2\eta}{3\gamma^2}R^{AB}_5(\omega_5)\wedge u^{C}\wedge u^{D}\wedge u^{E}\nn\\
& \qquad\qquad\qquad\quad~~
+\tfrac{1}{5\gamma^4}u^{A}\wedge u^{B}\wedge u^{C}\wedge u^{D}\wedge u^{E}\Big)\,,
\end{align}
which corresponds to a Lanczos-Lovelock Lagrangian in $D=5$ 
when we interpret $u^{A}$ as the f\"{u}nfbein.

As already mentioned earlier, the IW contraction $\gamma\rightarrow\infty$ 
of the 4-dimensional conformal algebra results in the Poincar\'{e} algebra. 
Accordingly, taking $\gamma\rightarrow\infty$ in the 
action~\eqref{eq:csforso42} yields precisely the Chern-Simons action for the 5-dimensional 
Poincar\'{e} group~\cite{banados1996higher}
\begin{equation}
S_{\text{CS}_5}[\omega_5,u_5]\Big|_{\gamma\rightarrow\infty}
=\tfrac{1}{4}\int_{M_5}\epsilon_{ABCDE}R^{AB}_5(\omega_5)\wedge R^{CD}_5(\omega_5)\wedge u^{E}\,.
\end{equation}

\subsection{4-covariant basis}\label{sec:csiso3+24+13+1likebasis}
We now further decompose the indices into $A=(a,5)$.
The connection in (\ref{connectionsplit}) further splits up as follows in the 4-covariant basis,
\begin{equation}\label{eq:connection2}
\bm{A}=\bm{\omega}+\bm{s}+\bm{u}+\bm{\mu}
=\tfrac{1}{2}\omega^{ab}\bm{J}_{ab}+s^{a}\bm{B}_{a}+u^{a}\bm{T}_{a}+\mu\bm{D}\,,
\end{equation}
where we denote $\bm{\omega}_4=\bm{\omega}$ and $\bm{u}_4=\bm{u}$. 
Defining $u^{5}\equiv\mu$, we obtain the action in terms of 4-covariant objects,
\begin{align}\label{eq:csomegashmu}
\nn &S_{\text{CS}_5}[\omega,s,u,\mu]\\
\nn &=\tfrac{1}{4}\int_{M_5}\epsilon_{abcd}\Big(R^{ab}(\omega)\wedge R^{cd}(\omega)\wedge\mu+2\eta\, R^{ab}(\omega)\wedge s^{c}\wedge s^{d}\wedge\mu+s^{a}\wedge s^{b}\wedge s^{c}\wedge s^{d}\wedge\mu\\
\nn &-4\,R^{ab}(\omega)\wedge T^{c}(\omega,s)\wedge u^{d}-4\eta\,T^{a}(\omega,s)\wedge s^{b}\wedge s^{c}\wedge u^{d}+\tfrac{4\eta}{3\gamma^2}\,T^{a}(\omega,s)\wedge u^{b}\wedge u^{c}\wedge u^{d}\\
 &-\tfrac{2\eta}{\gamma^2}\,R^{ab}(\omega)\wedge u^{c}\wedge u^{d}\wedge\mu-\tfrac{2}{\gamma^2}\,s^{a}\wedge s^{b}\wedge u^{c}\wedge u^{d}\wedge\mu+\tfrac{1}{\gamma^4}\, u^{a}\wedge u^{b}\wedge u^{c}\wedge u^{d}\wedge\mu\Big).
\end{align}
Here we have introduced the torsion 2-form,\footnote{Note that this is a slight abuse
of terminology since the genuine 4-dimensional torsion 2-form corresponds to
the sub-components $T^{a}_{\mu\nu}$, $\mu,\nu=0,\hdots3$. 
Similarly, we will sometimes refer to the 5-dimensional 1-form components
$`e^a_m`$ as the vielbein (even though the genuine vierbein is 
$`e^a_\mu`$).} $T^{a}(\omega,s)\equiv \d s^a+`\omega^a_b`\wedge s^b$. 
In section~\ref{sec:analysis4cov} we will study two different dimensional reduction schemes for which the
action (\ref{eq:csomegashmu}) becomes related to Einstein-Cartan and conformal gravity, respectively.

\subsection{4-covariant canonical basis}
In the 4-covariant canonical basis, the gauge connection splits up as follows,
\begin{equation}\label{connsu}
\bm{A}=\bm{\omega}+\bm{e}+\bm{h}+\bm{\mu}=
\tfrac{1}{2}\omega^{ab}\bm{J}_{ab}+e^{a}\bm{P}_{a}+h^{a}\bm{K}_{a}+\mu\bm{D}\,.
\end{equation}
The introduction of the matrix $M$ corresponds to a change of variables
in the action. The vierbeine $u^{a}$ and $s^{a}$ are exchanged against
two linear combinations, $e^{a}$ and $h^{a}$, defined via
\begin{equation}
\left(\begin{array}{c}
s^{a}\\
u^{a}
\end{array}
\right)=\left(\begin{array}{cc}
a\ & c\\
b\ & d
\end{array}
\right)\left(\begin{array}{c}
e^{a}\\
h^{a}
\end{array}
\right)=M^{\operatorname{T}}\left(\begin{array}{c}
e^{a}\\
h^{a}
\end{array}
\right)\,.
\label{eq:sofeh}
\end{equation}
The expression for the action is a bit lengthy and we provide it in equation (\ref{eq:thebigaction}).
Its interaction potential contains all possible wedge products
between $e^{a}$ and $h^{a}$ (each with an additional $\wedge \mu$) and is therefore reminiscent of 
ghost-free bimetric gravity in $D=4$.
The action is however not truly bimetric-like\footnote{The term
\textit{bimetric-like} is supposed to emphasize that the fields are not genuinely
4-dimensional objects, 
since, for example, $e^{a}=`e^{a}_m`(X)\d X^m$. 
After choosing a dimensional reduction scheme we will denote the genuinely 
4-dimensional fields with a bar, 
e.g.~$\bar{e}^{a}(x)=`e^{a}_\mu`(x)\d x^\mu$.}
because it contains only one spin connection $`\omega^{ab}_m`(X)\,\d X^m$.

Let us discuss two examples for actions in the canonical basis.

\subsubsection{Conformal basis}

When the parameters of the algebra are chosen as in~\eqref{eq:case1}, the 
action \eqref{eq:thebigaction} becomes
\begin{align}\label{eq:csgamma0case1}  
S_{\text{CS}_5}[\omega,e,h,\mu]&=\tfrac{1}{4}\int_{M_5}\epsilon_{abcd}
\Big(R^{ab}(\omega)\wedge R^{cd}(\omega)+8ac\eta\,R^{ab}(\omega)\wedge e^{c}\wedge h^{d}
\nn \\
&\qquad\qquad\qquad
+16a^2c^2\,e^{a}\wedge e^{b}\wedge h^{c}\wedge h^{d}\Big)\wedge\mu
+\gamma\, S_{\text{torsion}}[\omega,e,h]\,.
\end{align}
The precise form of $S_{\text{torsion}}$ is provided in equation (\ref{torsionterms}).
It contains torsion terms for the vielbeine for $e^{a}$ and $h^{a}$, 
which are required for making the theory $\SO(4,2)$ gauge invariant. 
These torsion terms are proportional to the parameter $\gamma$
and we can make them disappear by setting $\gamma=0$. This however
breaks the gauge symmetry since some of the commutation relations \eqref{eq:confalg} 
become ill-defined in this limit.

\subsubsection{Orthogonal basis}

If we choose the parameters corresponding to the orthogonal basis, c.f.~\eqref{eq:case2},
the general action \eqref{eq:thebigaction} becomes
\begin{align}\label{eq:csgamma0case3} 
\nn &S_{\text{CS}_5}[\omega,e,h,\mu]\\
\nn &=\tfrac{1}{4}\int_{M_5}\epsilon_{abcd}\Big(R^{ab}(\omega)\wedge R^{cd}(\omega)
+4a^2\eta\,R^{ab}(\omega)\wedge e^{c}\wedge e^{d}+4c^2\eta\,R^{ab}(\omega)\wedge h^{c}\wedge h^{d}\\
\nn &+4a^4\,e^{a}\wedge e^{b}\wedge e^{c}\wedge e^{d}
+8a^2c^2\,e^{a}\wedge e^{b}\wedge h^{c}\wedge h^{d}
+4c^4\,h^{a}\wedge h^{b}\wedge h^{c}\wedge h^{d}\Big)\wedge\mu\\
&+\gamma\, S'_{\text{torsion}}[\omega,e,h]\,.
\end{align}
The feature of this action is that the torsional part is again proportional to the parameter $\gamma$, which is not the case for general parameters in
the action \eqref{eq:thebigaction}. The torsion terms $S'_{\text{torsion}}[\omega,e,h]$ are given in 
equation~\eqref{torsionterms2}. To arrive at the action \eqref{eq:csgamma0case3}
we have performed a rescaling, $\gamma\rightarrow i\gamma$, which is necessary for the 
reality of the action and changes the $\SO(1,1)$ symmetry of the gauge algebra to 
$\SO(2)$ (provided that $\gamma^2>0$).

Since all algebras $\operatorname{C}_{3+1}(M, \gamma)$ are isomorphic, the corresponding actions
in $D=5$ will all be related through linear field redefinitions. 
In order to see this explicitly in a simple example, set $\gamma=0$ and take
\begin{align}\label{eq:redefinition}
e^{a} &\rightarrow \frac{1}{\sqrt{2}a}\Big(a\,e^{a}+ic\,h^{a}\Big)\,,\qquad
h^{a} \rightarrow \frac{1}{\sqrt{2}c}\Big(a\,e^{a}-ic\,h^{a}\Big)\,,
\end{align}
in \eqref{eq:csgamma0case1}. The resulting action will be~\eqref{eq:csgamma0case3}.
This equivalence of the actions expressed in different bases will be broken 
by our dimensional reduction schemes and hence the 4-dimensional theories we can derive 
from actions with different values for the matrix $M$ will not be indistinguishable.

\section{Dimensional reductions} \label{sec:csdimred}

In the general 5-dimensional Chern-Simons theory the fields depend on all coordinates, 
\begin{equation}
\bm{A}=\bm{A}_{m}(X)\,\d X^m\,,\qquad m=1,\ldots,5\,.
\end{equation}
In our reduction schemes, we will decompose $X^m$ into $X^\mu=x^\mu$ and $X^5=w$, which
are the coordinates on a 4- and a 1-dimensional submanifold $M_4$ and $\Sigma$ of $M_5$, 
such that $M_5=M_4\ltimes\Sigma$. We will then restrict the field dependences on these 
coordinates in different ways such that we can integrate over the 5th dimension.

\subsection{Reduction in the 4-covariant basis}\label{sec:analysis4cov}

In the following we analyze the relation of the 5-dimensional theory defined by the gauge invariant 
action~\eqref{eq:csomegashmu} to Einstein-Cartan and conformal gravity in $D=4$.

\subsubsection{Recovering Einstein-Cartan} \label{sec:4dECac}
Ref.~\cite{banados1997dilaton} studied the relation between a Chern-Simons action 
for the 5-dimensional Poincar\'{e} group in $D=5$ and a dilatonic Gauss-Bonnet 
term in $D=4$. Establishing this connection required
a certain dimensional reduction scheme and projecting the equations of motion onto the 4-dimensional boundary. 
Here we will invoke a different scheme which reduces our setup to Einstein-Cartan gravity
in $D=4$ with torsional counter-terms but without the dilaton.

First we take the limit $\gamma\rightarrow\infty$,
contracting the gauge algebra to the 5-dimensional Poincar\'{e} groups. 
Our action (\ref{eq:csomegashmu}) in the 4-covariant basis becomes
\begin{align}\label{eq:5cspoincare4basis} 
\nn &S_{\text{CS}_5}[\omega,s,u,\mu]\\
\nn &=\tfrac{1}{4}\int_{M_5}\epsilon_{abcd}\Big(R^{ab}(\omega)\wedge R^{cd}(\omega)
+2\eta\, R^{ab}(\omega)\wedge s^{c}\wedge s^{d}
+s^{a}\wedge s^{b}\wedge s^{c}\wedge s^{d}\Big)\wedge\mu\\
&-\int_{M_5}\epsilon_{abcd}
\Big(R^{ab}(\omega)+\eta\,s^{a}\wedge s^{b}\Big)\wedge T^{c}(\omega,s)\wedge u^{d}.
\end{align}
This is a Chern-Simons gauge theory in $D=5$ for the group 
$\ISO(3,2)$ or $\ISO(4,1)$ (depending on the sign of $\eta$), 
expressed in the components of the 4-covariant basis. 
Note that both gauge groups contain $\ISO(3,1)$ as a subgroup.
The action is invariant under the gauge transformations in eq.~\eqref{eq:gtpc5d}.

In order to reduce the theory to $D=4$, 
we now restrict the fields and their dependence on the coordinates as follows,
\begin{subequations}\label{eq:firstscheme}
\begin{align}
\label{eq:firstscheme1} `\omega^{ab}_{m}`(X)\,\d X^{m}
&=`\omega^{ab}_\mu`(x)\,\d x^\mu=:\bar{\omega}^{ab}(x)\,,\\
\label{eq:firstscheme2} `s^{a}_{m}`(X)\,\d X^{m}&=`s^{a}_{\mu}`(x)\,\d x^{\mu}=:\bar{s}^{a}(x)\,,\\
\label{eq:firstscheme3} `u^{a}_{m}`(X)\,\d X^{m}&=`u^{a}_{m}`(w)\,\d X^{m}=:\bar{u}^{a}(w)\,,\\
\label{eq:firstscheme4} \mu_{m}(X)\,\d X^{m}&=\mu_{m}(w)\,\d X^{m}=:\bar{\mu}(w)\,,
\end{align}
\end{subequations}
where the bars indicate the fact that this restriction corresponds to a field configuration, in
which the entire $\ISO(3,2)$ or $\ISO(4,1)$  
symmetry is broken. We can view it has a gauge-fixed version of more general
configurations that are obtained from (\ref{eq:firstscheme}) by gauge transformations.
In other words, we restrict the $\ISO(3,2)$ or $\ISO(4,1)$ 
gauge connection $\bm{A}'$ in the CS action to have components of the form,
\begin{align}
{\omega'}^{ab}(X)&=\bar{\omega}^{ab}(x)+\delta{\omega}^{ab}(X)\,,
\qquad 
{s'}^{a}(X)=\bar{s}^{a}(x)+\delta{s}^{a}(X)\,,
\nn\\
{u'}^{a}(X)&=\bar{u}^{a}(w)+\delta{u}^{a}(X)\,,
\qquad~~
{\mu'}(X)=\bar{\mu}(w)+\delta{\mu}(X)\,,
\end{align}
where the gauge transformations $\delta{\omega}^{ab}(X)$, etc.,
are given in eq.~\eqref{eq:gtpc5d}. Then the restricted action will still have the full 
$\ISO(3,2)$ or $\ISO(4,1)$ symmetry, but can be dimensionally
reduced only in the gauge (\ref{eq:firstscheme}). In this sense, the dimensional reduction
breaks the entire $\ISO(3,2)$ or $\ISO(4,1)$ symmetry.\footnote{Indeed
it is easy to show that the gauge-fixing (\ref{eq:firstscheme}) leaves no residual gauge transformations.}
Note that this means that we even lost local Lorentz invariance in $D=4$, which will become
explicit in what follows.

With components restricted as in (\ref{eq:firstscheme}), 
$R^{ab}(\bar{\omega})$ coincides with the Lorentz curvature and $\bar{s}^{a}$ with the vierbein
of the 4-dimensional Cartan formalism. 
Plugging the restricted fields into the (A)dS
invariant action (\ref{eq:5cspoincare4basis}) gives
\begin{align}
\nn \bar{S}[\bar{\omega},\bar{s},\bar{u},\bar{\mu}]
&=\tfrac{\eta}{\ell^2}\int_\Sigma\bar{\mu}\int_{M_4}\epsilon_{abcd}
\Big(\tfrac{\eta}{2} R^{ab}(\bar\omega)\wedge R^{cd}(\bar\omega)+
R^{ab}(\bar{\omega})\wedge\bar{s}^{c}\wedge\bar{s}^{d}
+\tfrac{\eta}{\ell^2}\,\bar{s}^{a}\wedge\bar{s}^{b}\wedge\bar{s}^{c}\wedge\bar{s}^{d}\Big)\\
\label{eq:ecplustorsion}
&+\tfrac{\sqrt{2}}{\ell}\int_\Sigma\bar{u}^{a}\int_{M_4}\epsilon_{abcd}\,
\Big(R^{bc}(\bar{\omega})+\tfrac{2\eta}{\ell^2}
\,\bar{s}^{b}\wedge\bar{s}^{c}\Big)\wedge \bar{T}^{d}(\bar{\omega},\bar{s}).
\end{align}
To arrive at this form, we have rescaled $\bar{s}^{a}\rightarrow(\sqrt{2}/\ell)\bar{s}^{a}$, 
where $\ell$ is a constant. The reduced action involves the torsion 2-form 
$\bar{T}^{d}(\bar{\omega},\bar{s})=D_{\bar{\omega}} \bar{s}$ 
and it is a sum of products of separate integrals.
We can now integrate over the fifth dimension and define the constants $\kappa=\int_\Sigma\bar{\mu}$, $\phi^{a}=\int_\Sigma\bar{u}^{a}$.
This produces the following 4-dimensional action,
\begin{align}
\nn \bar{S}[\bar{\omega},\bar{s}]&=\tfrac{\kappa\eta}{\ell^2}\int_{M_4}\epsilon_{abcd}\Big(R^{ab}(\bar{\omega})\wedge\bar{s}^{c}\wedge\bar{s}^{d}+\tfrac{\eta}{\ell^2}\,\bar{s}^{a}\wedge\bar{s}^{b}\wedge\bar{s}^{c}\wedge\bar{s}^{d}\Big)\\
&+\tfrac{\sqrt{2}}{\ell}\phi^{a}\int_{M_4}\epsilon_{abcd}\,\Big(R^{bc}(\bar{\omega})+\tfrac{2\eta}{\ell^2}\,\bar{s}^{b}\wedge\bar{s}^{c}\Big)\wedge \bar{T}^{d}(\bar{\omega},\bar{s})\,.
\end{align}
Here we have omitted the topological Gauss-Bonnet term. 
The action corresponds precisely to the first-order 4-dimensional 
Einstein-Cartan theory with negative or positive cosmological constant for $\eta=\pm1$, respectively,
plus terms including torsion. 

Note that the presence of the constant vector $\phi^{a}$ breaks the local Lorentz symmetry.
The only ways to reinstore 4-dimensional Lorentz invariance are to assume vanishing torsion 
or to demand that $\phi^{a}=0$, which is a further strong restriction on the gauge field component $u^a$.
In both cases we recover precisely the Einstein-Cartan action.

\subsubsection{A first-order action for conformal gravity} \label{sec:recovCG}

We now go back to the $\SO(4,2)$ invariant action \eqref{eq:csomegashmu} with arbitrary $\gamma$
and perform a dimensional reduction. 
We restrict the gauge field components in a slightly different way,
\begin{subequations}\label{eq:secondscheme} 
\begin{align}
\label{eq:secondscheme1} `\omega^{ab}_m`(X)\,\d X^m&=`\omega^{ab}_\mu`(x)\,\d x^\mu=:\bar{\omega}^{ab}(x)\,,\\
\label{eq:secondscheme2} `s^{a}_m`(X)\,\d X^m&=`s^{a}_\mu`(x)\,\d x^\mu=:\bar{s}^{a}(x)\,,\\
\label{eq:secondscheme3} `u^{a}_m`(X)\,\d X^m&=`u^{a}_\mu`(x)\,\d x^\mu=:\bar{u}^{a}(x)\,,\\
\label{eq:secondscheme4} \mu_m(X)\,\d X^m&=\mu_5(w)\,\d w=:\bar{\mu}(w)\,.
\end{align}
\end{subequations}
In this case the semi-direct product $M_5=M_4\ltimes\Sigma$ becomes a direct product $M_5=M_4\times\Sigma$
since the components of the f\"{u}nfbein $`s^{A}_{m}`(X)$ satisfy 
$`s^{a}_{5}`=`s^{5}_{\mu}`=0$ and the same for $`u^A_m`(X)$.

Analogously to the scheme in the previous subsection, these can be viewed as a gauge-fixed version of more
general configurations which can be obtained from (\ref{eq:secondscheme}) through
the general $\SO(4,2)$ transformations displayed in~\eqref{eq:gauget4covb}. 
However, in this case, \eqref{eq:secondscheme} do not fix the gauge completely. 
One can still perform transformations with parameters 
$\theta^{ab}=\theta^{ab}(x)$ which appear in~\eqref{eq:gauget4covb} for $\beta^a=\tau^a=\lambda=0$.
These correspond to a residual symmetry which is precisely
the 4-dimensional local Lorentz symmetry $\SO(3,1)$.

The above restriction $\mu=\mu_5\,\d w$ ensures that all other components parallel to $\d w$, 
including all torsion terms, will drop out from the action due to $\d w\wedge\d w=0$.
Thus all appearing indices on the other fields will be $D=4$ spacetime indices.
The action \eqref{eq:csomegashmu} reduces to
\begin{align}
\nn \bar{S}[\bar{\omega},\bar{s},\bar{u},\bar{\mu}]
&=\tfrac{\eta}{2}\int_\Sigma\bar{\mu}\int_{M_4}\epsilon_{abcd}
\Big(\tfrac{\eta}{2}R^{ab}(\bar{\omega})\wedge R^{cd}(\bar{\omega})+
R^{ab}(\bar{\omega})\wedge\Big[\bar{s}^{c}\wedge\bar{s}^{d}
-\tfrac{1}{\gamma^2}\bar{u}^{c}\wedge\bar{u}^{d}\Big]\\
&\qquad \qquad\qquad
+\tfrac{\eta}{2}\Big[\bar{s}^{a}\wedge\bar{s}^{b}
-\tfrac{1}{\gamma^2}\bar{u}^{a}\wedge\bar{u}^{b}\Big]\wedge
\Big[\bar{s}^{c}\wedge\bar{s}^{d}-\tfrac{1}{\gamma^2}\bar{u}^{c}\wedge\bar{u}^{d}\Big]\Big).
\end{align}
Defining the constant
$
\kappa=\int_\Sigma\bar{\mu}
$
the action in $D=4$ becomes
\begin{align}\label{cgdr}
\nn \bar{S}[\bar{\omega},\bar{s},\bar{u}]&=
\tfrac{\kappa\eta}{2}\int_{M_4}\epsilon_{abcd}
\Big(R^{ab}(\bar{\omega})\wedge\Big[\bar{s}^{c}\wedge\bar{s}^{d}
-\tfrac{1}{\gamma^2}\bar{u}^{c}\wedge\bar{u}^{d}\Big]\\
&\qquad \qquad 
+\tfrac{\eta}{2}\Big[\bar{s}^{a}\wedge\bar{s}^{b}
-\tfrac{1}{\gamma^2}\bar{u}^{a}\wedge\bar{u}^{b}\Big]\wedge
\Big[\bar{s}^{c}\wedge\bar{s}^{d}-\tfrac{1}{\gamma^2}\bar{u}^{c}\wedge\bar{u}^{d}\Big]\Big)\,,
\end{align}
where we have again omitted the Gauss-Bonnet term.
This result looks very similar to the conformal gravity action in~\eqref{eq:focg}. 
However, the crucial difference is that the spin connection is an independent field in
our setup and hence the action is not equivalent to~\eqref{eq:focg} where $\bar{\omega}$ 
is fixed to be the spin connection for the vierbein $s^a+\gamma^{-1}u^a$.
The invariance under Weyl rotations 
of the vierbein vector $(u^a, \gamma s^a)$ is a direct
consequence of the $\SO(1,1)$ symmetry of the $\SO(4,2)$ algebra, c.f.~equation~(\ref{algrot}).

The action (\ref{cgdr}) is precisely the same as the one obtained in Ref.~\cite{kaku1977gauge} which  
constructed a 4-dimensional gauge theory in terms of the unique quadratic action 
for the curvatures of $\SO(4,2)$.
Our result differs from Ref.~\cite{aros2014ads} which derived conformal gravity
from a 5-dimensional gauge theory for the group $\SO(4,2)$. 
For performing the dimensional reduction, this reference assumed the existence of an isometry along the fifth 
dimension such that no fields depended on it.
In this way they obtained conformal gravity with an additional dilatonic field under the assumption of
vanishing torsion. 

\subsection{Reduction in the orthogonal basis}

As stated earlier, the 5-dimensional theories expressed in different bases of the algebra
are all equivalent because they can be obtained from one to another by linear field redefinitions.
However, in this section we will introduce a dimensional reduction scheme which explicitly breaks this
equivalence and makes the 4-dimensional setup basis-dependent.
Namely, we will dimensionally reduce the theory
by restricting the gauge field components as follows,
\begin{subequations}\label{eq:dimreddcb}
\begin{align}
\label{eq:omdimred} `\omega^{ab}_{m}`(X)\,\d X^m
&=`\omega^{ab}_{\mu}`(x)\,\d x^\mu+`\omega^{ab}_{5}`(x,w)\,\d w
=:\bar{\omega}^{ab}(x)+`\omega^{ab}_{5}`(x,w)\,\d w\,,\\
\label{eq:eansatz} `e^{a}_{m}`(X)\,\d X^m
&=e(w)\,`e^{a}_{\mu}`(x)\,\d x^\mu+`e^{a}_{5}`(x,w)\,\d w
=:e(w)\,\bar{e}^{a}(x)+`e^{a}_{5}`(x,w)\,\d w\,,\\
\label{eq:hansatz} `h^{a}_{m}`(X)\,\d X^m
&=h(w)\,`h^{a}_{\mu}`(x)\,\d x^\mu+`h^{a}_{5}`(x,w)\,\d w
=:h(w)\,\bar{h}^{a}(x)+`h^{a}_{5}`(x,w)\,\d w\,,\\
\label{eq:mudimred} \mu_m(X)\,\d X^m
&=\mu_5(w)\,\d w
=:\bar{\mu}(w)\,.
\end{align}
\end{subequations}
Here we have introduced two arbitrary scalar functions $e=e(w)$ and $h=h(w)$ on $\Sigma$. 
They define a warped spacetime in the direction $X^5=w$, since
the metric of a warped geometry has the general form,
\begin{equation}\label{eq:geomin}
\d s^2=f(w)g_{\mu\nu}(x)\d x^\mu\d x^\nu+g(w)\d w^2\,.
\end{equation}
For a discussion of effective theories with warped geometry see, for example, 
Ref.~\cite{Izaurieta:2012fi}. 

The form of the restriction (\ref{eq:dimreddcb})
is not the same for linear combinations of the 5-dimensional fields $e^a$ and $h^a$
due to the presence of the warp functions.\footnote{More precisely, a linear combination
$l^a=ae^a+bh^a$ is not writable in the form $l^a=l(w)\,\bar{l}^{a}(x)+`l^{a}_{5}`(x,w)\,\d w$
but it will read $l^a=ae(w)\,\bar{e}^{a}(x)+bh(w)\,\bar{h}^{a}(x)+`l^{a}_{5}`(x,w)\,\d w$.}
Thus, our dimensional reduction requires picking a particular basis of the gauge algebra. 
Clearly, this leaves us with infinitely many choices to reduce the Chern-Simons action to $D=4$,
corresponding to all possible different matrices $M$ that define $\operatorname{C}_{3+1}(M,\gamma)$. 

As an example, we choose here the orthogonal basis whose torsional part can be removed by
setting $\gamma=0$.
We then reduce the action (\ref{eq:csgamma0case3}) with the restriction (\ref{eq:dimreddcb}).
As in our previous examples, we can allow the fields in the action to be gauge transformations
of (\ref{eq:dimreddcb}). The transformations in the orthogonal basis are provided in
\eqref{eq:gtdcb}. It turns out that (\ref{eq:dimreddcb}) does not completely fix the gauge
and there are two possibilities for residual gauge transformations:\\
{\bf i)} A transformation with $\theta^{ab}=\theta^{ab}(x)$ corresponding to 4-dimensional 
 local Lorentz symmetry. In this case, the functions $e(w)$ and $h(w)$ remain arbitrary.\\
{\bf ii)} $\theta^{ab}=\theta^{ab}(x)$ and $\lambda=\lambda(w)$ corresponding to 4-dimensional 
local Lorentz symmetry plus a Weyl symmetry in the fifth dimension.
In this case, the functions $e(w)$ and $h(w)$ are required to be proportional. 

In both cases the residual gauge transformations do not change the functions 
$e(w)$ and $h(w)$. Since we are interested in the most general 4-dimensional action,
in the following we will restrict ourselves to the possibility i) 
i.e.~we will have 4-dimensional local Lorentz symmetry and the functions $e(w)$ and $h(w)$ 
remain unrestricted.

With the fields restricted to satisfy \eqref{eq:dimreddcb},  
the action \eqref{eq:csgamma0case3} for $\gamma=0$ reduces to
\begin{align}
\nn &\bar{S}[\bar\omega,\bar{e},\bar{h},\bar{\mu}]
\nn =a^2\eta\int_\Sigma e^2\bar\mu\int_{M_4}\epsilon_{abcd}\,R^{ab}(\bar{\omega})\wedge\bar{e}^{c}\wedge\bar{e}^{d}+c^2\eta\int_\Sigma h^2\bar\mu\int_{M_4}\epsilon_{abcd}\,R^{ab}(\bar{\omega})\wedge\bar{h}^{c}\wedge\bar{h}^{d}\\
\nn &+a^4\int_\Sigma e^4\bar\mu\int_{M_4}\epsilon_{abcd}\,\bar{e}^{a}\wedge\bar{e}^{b}\wedge\bar{e}^{c}\wedge\bar{e}^{d}+2a^2c^2\int_\Sigma e^2h^2\bar\mu\int_{M_4}\epsilon_{abcd}\,\bar{e}^{a}\wedge\bar{e}^{b}\wedge\bar{h}^{c}\wedge\bar{h}^{d}\\
&+c^4\int_\Sigma h^4\bar\mu\int_{M_4}\epsilon_{abcd}\,\bar{h}^{a}\wedge\bar{h}^{b}\wedge\bar{h}^{c}\wedge\bar{h}^{d}\,,
\end{align}
where $e^2$, $h^2$, etc.~are powers of the warp functions. We emphasize again
that, since we took $\gamma\rightarrow0$, the $\operatorname{C}_{3+1}(M,\gamma)$ algebra 
is undefined and we have lost the gauge invariance. Nevertheless, the 4-dimensional theory
still has a residual local Lorentz invariance.

To integrate over the fifth dimension, we define the constants
\begin{equation}\label{eq:pquantities} 
p_{st}=\int_\Sigma\d w\, e^s(w)h^t(w)\mu_5(w)\,,
\end{equation}
which involve different powers of the warping functions.
The 4-dimensional action becomes
\begin{align}\label{eq:csgamma0case3dimred} 
\nn &\bar{S}[\bar\omega,\bar{e},\bar{h}]
=\int_{M_4}\epsilon_{abcd}\Big(a^2\eta p_{20}\,R^{ab}(\bar{\omega})\wedge\bar{e}^{c}\wedge\bar{e}^{d}
+c^2\eta p_{02}\,R^{ab}(\bar{\omega})\wedge\bar{h}^{c}\wedge\bar{h}^{d}\\
&+a^4 p_{40}\,\bar{e}^{a}\wedge\bar{e}^{b}\wedge\bar{e}^{c}\wedge\bar{e}^{d}
+2a^2c^2p_{22}\,\bar{e}^{a}\wedge\bar{e}^{b}\wedge\bar{h}^{c}\wedge\bar{h}^{d}
+c^4p_{04}\,\bar{h}^{a}\wedge\bar{h}^{b}\wedge\bar{h}^{c}\wedge\bar{h}^{d}\Big)\,.
\end{align}
For a special relation among the parameters $p_{st}$, 
\begin{equation}
p_{40}=p_{20}^2\,,\qquad p_{22}=p_{20}p_{02}\,,\qquad p_{04}=p_{02}^2\,,
\label{eq:psconf}
\end{equation}
we perform the following field redefinition,
\begin{align}\label{fredi}
a'E^{a}&= a\sqrt{\eta p_{20}}\,\bar{e}^{a}+ic\sqrt{\eta p_{02}}\,\bar{h}^{a}\,,\qquad
c'H^{a}=a\sqrt{\eta p_{20}}\,\bar{e}^{a}-ic\sqrt{\eta p_{02}}\,\bar{h}^{a}\,.
\end{align}
In terms of these new variables, the action assumes the form of first-order conformal gravity,
just like in the 4-covariant basis, c.f.~section~\ref{sec:recovCG}.
For more general parameter choices, the action (\ref{eq:csgamma0case3dimred}) can be viewed as a generalization
of first-order conformal gravity which is no longer invariant under the $\SO(2)$ rotations. 
In this case the Weyl rotation invariance is explicitly broken by an asymmetric choice of parameters $p_{st}$
which implies an asymmetry in the warp functions $e(w)$ and $h(w)$ for the two vierbeine.

\subsection{Generalized bimetric gravity from a doubled Chern-Simons theory}

Ref.~\cite{Rahmanpour:2018lzl} derived a generalized massive gravity model with additional fields from
a 4-dimensional gauge theory of $\SO(4,2)$. Ref.~\cite{apolo2017non} started from the group $\SO(5,1)$
to arrive at a model that shows similarities to both bimetric and conformal gravity but contains an 
additional vector field. Here we will show how bimetric interactions (in vierbein
formulation) can emerge from 5-dimensional Chern-Simons terms.

To this end, we now consider a doubled version of the Chern-Simons theory in the 4-covariant canonical basis,
by adding another gauge invariant action ${S}_{\mathrm{CS}_5}[\tilde{\omega},e,h,\tilde{\mu}]$
to \eqref{eq:csgamma0case3}.\footnote{In spirit, our doubled Chern-Simons action is somewhat similar 
to theories built from transgression forms~\cite{Zanelli:2005sa, Mora:2006ka}.} 
 The second copy involves a different spin connection $\tilde{\omega}$ 
and a different one-form $\tilde{\mu}$, whereas the two vierbein fields are taken to be the same as in 
${S}_{\mathrm{CS}_5}[{\omega},e,h,{\mu}]$. Due to its interdependent field content, the sum 
${S}_{\mathrm{CS}_5}[{\omega},e,h,{\mu}]+ {S}_{\mathrm{CS}_5}[\tilde{\omega},e,h,\tilde{\mu}]$
breaks the $\SO(4,2)\times\SO(4,2)$ gauge symmetry to a smaller subgroup.
As can be read off from the gauge transformations in equation \eqref{eq:gauget4covcanb}, 
the residual gauge symmetry of the action is 
$\SO(3,1)\times\text{Dilation}$ with gauge parameters $\theta^{ab}$ and $\lambda$.

We then perform a dimensional reduction
by restricting both sets of fields as in \eqref{eq:dimreddcb} 
with analogous forms for $\tilde{\omega}`{}^ab_m`$ and $\tilde{\mu}_m$ . The action reduces to,
\begin{align}\label{eq:bmgen} 
\nn &\overline{S+\tilde{S}}
=\eta\int_{M_4}\epsilon_{abcd}\Big(a^2 p_{20}\,R^{ab}(\bar{\omega})\wedge\bar{e}^{c}\wedge\bar{e}^{d}
+c^2 p_{02}\,R^{ab}(\bar{\omega})\wedge\bar{h}^{c}\wedge\bar{h}^{d}\\
\nn  &\qquad\qquad\qquad\qquad\qquad\qquad\qquad\qquad
+a^2 \tilde{p}_{20}\,R^{ab}(\bar{\tilde{\omega}})\wedge\bar{e}^{c}\wedge\bar{e}^{d}
+c^2 \tilde{p}_{02}\,R^{ab}(\bar{\tilde{\omega}})\wedge\bar{h}^{c}\wedge\bar{h}^{d}\Big)\\
\nn &\qquad\qquad
+\int_{M_4}\epsilon_{abcd}\Big(a^4(p_{40}+\tilde{p}_{40})
\,\bar{e}^{a}\wedge\bar{e}^{b}\wedge\bar{e}^{c}\wedge\bar{e}^{d}
+2a^2c^2(p_{22}+\tilde{p}_{22})\,\bar{e}^{a}\wedge\bar{e}^{b}\wedge\bar{h}^{c}\wedge\bar{h}^{d}\\
&\qquad\qquad\qquad\qquad\qquad\qquad\qquad\qquad
+c^4(p_{04}+\tilde{p_{04}})
\,\bar{h}^{a}\wedge\bar{h}^{b}\wedge\bar{h}^{c}\wedge\bar{h}^{d}\Big)+ 
\gamma\bar{S}_{\text{torsion}}\,.
\end{align}
where the bar over $S+\tilde{S}$ is to make clear that we dimensionally reduced the sum of 
both Chern-Simons gauge theories and $\bar{S}_{\text{torsion}}$ is 
the dimensional reduced version of the sum of the two torsion terms 
$S'_{\text{torsion}}+\tilde{S}'_{\text{torsion}}$. 
We have defined the constants,
\begin{equation}\label{eq:ptildequantities} 
\tilde{p}_{st}=\int_\Sigma\d w\, e^s(w)h^t(w)\tilde{\mu}_5(w)\,.
\end{equation}
For general values of these constants,
the above restrictions on the fields and reduction to $D=4$ further breaks the gauge symmetry
down to $\SO(3,1)$.
The resulting action in eq.~\eqref{eq:bmgen} can be seen as a generalization of bimetric theory 
with interaction parameters $\beta_1=\beta_3=0$. 

We now set $\gamma=0$, making the algebra ill-defined, and 
identify two special cases:
\begin{itemize}
\item {\bf Bimetric gravity:} 
Since the $p$'s and $\tilde{p}$'s are all independent parameters from the 4-dimensional point of view, 
the potential in \eqref{eq:bmgen} is a special case of ghost-free bimetric theory. 
The kinetic terms are all accompanied by arbitrary parameters which we can 
restrict in order to guarantee the absence of ghosts.
Namely, the parameter choice
\beqn
p_{02}=0\,,\qquad\tilde{p}_{20}&=0
\,,
\eeqn
eliminates the kinetic mixings. In this case we can rename
\begin{align}
\eta a^2 p_{20}=\tfrac{1}{2}m_e^2\,,\qquad
\eta c^2\tilde{p}_{02}=\tfrac{1}{2}m_h^2\,,\qquad
a^4(p_{40}+\tilde{p}_{40})&=-\tfrac{1}{8\cdot4!}m^4\beta_0\,,
\nonumber
\\
2a^2c^2(p_{22}+\tilde{p}_{22})=-\tfrac{1}{8\cdot 4}m^4\beta_2\,,\qquad
c^4(p_{04}+\tilde{p_{04}})&=-\tfrac{1}{8\cdot 4!}m^4\beta_4\,,
\end{align}
such that the action \eqref{eq:bmgen} becomes (for $\gamma=0$)
\begin{align}
\nn S&=\frac{1}{2}\int_{M_4}\epsilon_{abcd}
\Big(m_e^2\,R^{ab}(\bar{\omega})\wedge\bar{e}^{c}\wedge\bar{e}^{d}
+m_h^2\,R^{ab}(\bar{\tilde{\omega}})\wedge\bar{h}^{c}\wedge\bar{h}^{d}\Big)\\
\label{eq:b1b3bg} &-\frac{m^4}{8}\int_{M_4}\epsilon_{abcd}
\Big(\tfrac{\beta_0}{4!}\,\bar{e}^{a}\wedge\bar{e}^{b}\wedge\bar{e}^{c}\wedge\bar{e}^{d}
+\tfrac{\beta_2}{4}\,\bar{e}^{a}\wedge\bar{e}^{b}\wedge\bar{h}^{c}\wedge\bar{h}^{d}
+\tfrac{\beta_4}{4!}\,\bar{h}^{a}\wedge\bar{h}^{b}\wedge\bar{h}^{c}\wedge\bar{h}^{d}\Big)\,.
\end{align}
This is bimetric gravity in first-order formulation with $\beta_1=\beta_3=0$. 
It is invariant under the residual local Lorentz symmetry, as expected.
We leave it as an open question whether the action with more general parameters
propagates the Boulware-Deser ghost.
Note also that the special model of eq.~(\ref{genact}) corresponds to taking
\begin{equation}
p_{40}+\tilde{p}_{40}=p_{20}^2\,,\quad p_{22}+\tilde{p}_{22}
=p_{20}\tilde{p}_{02},\quad p_{04}+\tilde{p}_{04}=\tilde{p}_{02}^2\,
\quad \tilde{p}_{20}=0\,,
\quad p_{02}=0\,.
\label{eq:coeffforpm}
\end{equation}

\item {\bf Weyl rotation symmetry:} As we pointed out in section~\ref{sec:cgc}, 
the Weyl rotation symmetry of first-order conformal gravity is a subgroup of the conformal group. 
In fact there is a parameter choice that maintains the Weyl rotation symmetry of the 
action \eqref{eq:bmgen}. Namely, if instead of \eqref{eq:coeffforpm}, 
we choose the parameters as,
\beqn
p_{40}+\tilde{p}_{40}=p_{20}^2\,,\ p_{22}+\tilde{p}_{22}=p_{20}\tilde{p}_{02}\,,\ p_{04}+\tilde{p}_{04}=\tilde{p}_{02}^2\,,\ p_{02}=\tilde{p}_{02}\,,\ \tilde{p}_{20}=p_{20}\,,
\eeqn
the action \eqref{eq:bmgen} for $\gamma=0$ takes the form
\begin{align}
\nn &S
=\eta\int_{M_4}\epsilon_{abcd}\Big(\Big[R^{ab}(\bar{\omega})+R^{ab}(\bar{\tilde{\omega}})\Big]\wedge\Big[a^2 p_{20}\,\bar{e}^{c}\wedge\bar{e}^{d}
+c^2 \tilde{p}_{02}\,\bar{h}^{c}\wedge\bar{h}^{d}\Big]\\
&\qquad\qquad+\Big[a^2 p_{20}\,\bar{e}^{a}\wedge\bar{e}^{b}
+c^2 \tilde{p}_{02}\,\bar{h}^{a}\wedge\bar{h}^{b}\Big]\wedge\Big[a^2 p_{20}\,\bar{e}^{c}\wedge\bar{e}^{d}
+c^2 \tilde{p}_{02}\,\bar{h}^{c}\wedge\bar{h}^{d}\Big]\Big)\,.
\end{align}
This action is invariant under the $\SO(2)$ 
(or $\SO(1,1)$, depending on the sign of $p_{20}\tilde{p}_{02}$) Weyl rotation invariance. 
Note, however, that it differs from first-order conformal gravity due to the appearance of two curvature
forms depending on two independent spin connections.

\end{itemize}

\section{Discussion}\label{sec:discussion}

We showed how an extended version of Einstein-Cartan theory in $D=4$ 
can be obtained from a 5-dimensional Chern-Simons action for the gauge group $\SO(4,2)$ after an IW contraction.
Away from this contraction limit, the 4-dimensional theory is a first-order version of conformal gravity.
Parametrizing the gauge algebra of $\SO(4,2)$ in terms of the matrix $M$ and the IW parameter $\gamma$ then allowed
us to construct a canonical basis of generators and perform further inequivalent dimensional reductions
including warp functions. 
In this way we discovered new connections between spin-2 field theories in $D=4$ and Chern-Simons geometries
in $D=5$. In particular, we arrived at a family of generalized conformal gravity actions.
Finally, invoking two copies of the Chern-Simons action with interdependent field content, 
dimensional reduction resulted in a family of generalized bimetric theories which contain the ghost-free 
theory as a subclass.

Let us briefly comment on how our approach differs from earlier 
work on dimensional reductions of the Chern-Simons action. 
Ref.~\cite{chamseddine1990topological} reduced the $\SO(4,2)$ 
Chern-Simons theory to get the so-called topological gravity. 
The result is different from ours due to different restrictions on the vierbein components and
on the field dependence on the 5th coordinate. 
The author of Ref.~\cite{banados1997dilaton} performed a projection of $\ISO(4,1)$ and $\SO(4,2)$ 
Chern-Simons theories onto $D=4$ at the level of equations of motion. 
In the $\SO(4,2)$ case this resulted in standard GR plus a dilatonic Gauss-Bonnet term. 
In Ref.~\cite{Aros:2007nn} the same Chern-Simons theories
are dimensional reduced using a Kaluza-Klein ansatz and demanding vanishing torsion. 
A similar approach was taken in Ref.~\cite{Izaurieta:2012fi} with the difference 
that their ansatz is a warp geometry instead of Kaluza-Klein. The procedure in our paper is yet a different 
one since we do not restrict to $T^a=0$ and impose different conditions
on the fields in our dimensional reduction schemes. 
Further models with vanishing torsion and a diagonal metric ansatz for the 
5-dimensional geometry were studied in Ref.~\cite{Morales:2016nva, Morales:2017zjw}.
In Ref.~\cite{aros2014ads} the Chern-Simons theory for $\SO(4,2)$ 
is dimensional reduced in a scheme with vanishing torsion and a particular tractor connection.
The resulting effective Lagrangian is standard conformal gravity multiplied by a scalar function.

Our dimensional reductions are imposed as external restrictions on the fields; we have not checked whether
they can arise dynamically as, for instance, in Ref.~\cite{Nastase:2007ma}. 
Moreover, it is important to note that the procedure of dimensional reduction with warp 
functions breaks the equivalence of the 5-dimensional actions
and we have only studied some examples for the resulting 4-dimensional field theories. 
We thus emphasize that all our results depend on choices for the dimensional reduction scheme which introduces some 
level of arbitrariness into the setup. This is reflected by the fact that earlier results in the literature
included many different actions in $D=4$ derived from the same 5-dimensional theory.
Nevertheless, our analysis demonstrates that a possible origin for structures
of 4-dimensional spin-2 interactions (which have not been obtained before)
could be found in 5-dimensional Chern-Simons theories. 

The dimensional reduction breaks the gauge group $\SO(4,2)$ to a smaller subgroup, which corresponds
to the Lorentz invariance in $D=4$ (except for the case of Einstein-Cartan gravity where without
further restrictions no residual symmetries survive in $D=4$).
An interesting observation is that the Weyl rotation invariance of first-order conformal gravity
(which was already present in the purely 4-dimensional setup of Ref.~\cite{kaku1977gauge}) 
stems from an $\SO(1,1)$ invariance of the conformal algebra. Introducing the warp functions
into the dimensional reduction generically explicitly breaks this symmetry. To our knowledge, 
the resulting generalizations of first-order conformal gravity and bimetric theory have not been 
discussed in the literature before and it would be interesting to study their properties. In 
particular, an important question is whether they are free from ghost instabilities.


\begin{acknowledgments}
NLGA would like to thank Patricio Salgado, Michael Haack, Matteo Parisi, Andrea Orta, Fabrizio Cordonier-Tello, Emmanuel Malek, Gustavo Rubio, Eduardo Rodr\'{i}guez, Enrico Andriolo and George Zoupanos for enlightening discussions. This research was partially funded by the bilateral DAAD-CONICYT grant 72150534 (NLGA) and 62160015 (SS). The work of DL is supported by the ERC Advanced Grant ''Strings and Gravity`` (Grant No. 320045) and the Excellence Cluster Universe. ASM acknowledges support from the Max-Planck-Society.
\end{acknowledgments}


\appendix


\section{Gauge theory for the conformal group}\label{sec:gaugethc}

In this appendix we summarize aspects of the gauge theory for the group 
$\SO(4,2)$ in the different bases of in section~\ref{sec:confalg}. 
We use the following conventions for coordinate indices,
\begin{align*}
\mu,\nu,\ldots&=1,2,3,4\,,\\
m,n,\ldots&=1,2,3,4,5\,,
\end{align*}
and internal indices,
\begin{align*}
a,b,\ldots&=1,2,3,4\,,\\
A,B,\ldots&=1,2,3,4,5\,,\\
I,J,\ldots&=1,2,3,4,5,6\,.
\end{align*}

\subsection{Connections and curvatures}

We begin by summarizing the notation for the spin connections 
and curvatures appearing in this manuscript.
\begin{itemize}
\item The conformal spin connection on $(M_6,\SO(4,2))$ is defined as
\begin{align}
\bm{A}(X,X^6)=\frac{1}{2}A^{IJ}(X,X^6)\bm{J}_{IJ}\,,\qquad 
\operatorname{span}\set{\bm{J}_{IJ}}=\SO(4,2)\,.
\end{align}
The field strength $\bm{F}_{\bm{A}}=\d\bm{A}+\tfrac{1}{2}\comm{\bm{A}}{\bm{A}}$ 
is exactly the curvature of the Cartan formalism for the special orthogonal group. 
In the 6-covariant basis it is given by
\begin{equation}
\bm{F}_{\bm{A}}=\bm{R}_6(\bm A)=\tfrac{1}{2}R^{IJ}(A)\bm{J}_{IJ}
=\tfrac{1}{2}\Big(\d A^{IJ}+`A^I_K`\wedge A^{KJ}\Big)\bm{J}_{IJ}\,,
\end{equation}
where $R^{IJ}(A)$ is the conformal curvature.
The decomposition of the gauge connection into the 5-covariant basis reads
\begin{align}
\bm{A}=\tfrac{1}{2}{A}^{IJ}\bm{J}_{IJ}
=\bm{\omega}_5+\bm{u}_{5}=\tfrac{1}{2}\omega^{AB}\bm{J}_{AB}+u^{A}\bm{T}_{A}\,,
\end{align}
and into the 4-covariant basis it is given by
\begin{align}
\bm{A}=\bm{\omega}+\bm{s}+\bm{u}+\bm{\mu}
=\tfrac{1}{2}\omega^{ab}\bm{J}_{ab}+s^{a}\bm{B}_{a}+u^{a}\bm{T}_{a}+\mu\bm{D}\,.
\end{align}

\item The (A)dS spin connection on $(M_5,\SO(3,2))$ or 
$(M_5,\SO(4,1))$ is defined as
\begin{displaymath}
\bm{\omega}_5(X)=\tfrac{1}{2}\omega^{AB}(X)\bm{J}_{AB}\,,\qquad
\operatorname{span}\set{\bm{J}_{ab}}=\SO(3,2)~~\text{ or }~~\operatorname{span}\set{\bm{J}_{AB}}=\SO(4,1)\,.
\end{displaymath}
In the 5-covariant basis the field strength reads
\begin{align}\label{eq:sopq1likecurv5}
\bm{F}_{\bm{\omega}_{5}+\bm{u}_{5}}
&=\bm{R}_{5}(\bm\omega_5)+\tfrac{1}{2}\comm{\bm{u}_{5}}{\bm{u}_{5}}+\D_{\bm{\omega}_{5}}\bm{u}_{5}
\nn\\
&=\tfrac{1}{2}\Big(R^{AB}(\omega)-\eta\gamma^{-2}u^A\wedge u^B\Big)\bm{J}_{AB}
+\D_{\omega}u^A\,\bm{T}_A\,,
\end{align}
where
\begin{equation}
\bm{R}_5(\bm\omega_5)
=\tfrac{1}{2}R^{AB}(\omega)\bm{J}_{AB}\,,\qquad R^{AB}(\omega)
=\d\omega^{AB}+`\omega^A_C`\wedge\omega^{CB}\,,\\
\label{eq:sopq1likecurv1}
\end{equation}
is the (A)dS spin curvature and $\D_{\omega}u^A=\d u^A+`\omega^A_B`\wedge u^B$.

\item The Lorentz spin connection one-form on $(M_4,\SO(3,1))$ is defined as
\begin{equation}
\bm{\omega}(x):=\tfrac{1}{2}\omega^{ab}(x)\bm{J}_{ab}\,,\qquad
\operatorname{span}\set{\bm{J}_{ab}}=\SO(3,1)\,.
\end{equation}
The field strength in the 4-covariant basis is given by
\begin{align}\label{eq:sop+1q+1p+qlikecurvature2} 
\bm{F}_{\bm{\omega} +\bm{s}+\bm{u} +\bm{\mu}}
&=\bm{R}(\bm{\omega})+\D_{\bm{\omega}}\bm{s}
+\D_{\bm{\omega}}\bm{u}+\D_{\bm{\omega}}\bm{\mu}\nn \\
&~~~~~~~~~~~+\tfrac{1}{2}\comm{\bm{s}}{\bm{s}}+\comm{\bm{s}}{\bm{u}}+\comm{\bm{s}}{\bm{\mu}}
+\tfrac{1}{2}\comm{\bm{u}}{\bm{u}}+\comm{\bm{u}}{\bm{\mu}}+\tfrac{1}{2}\comm{\bm{\mu}}{\bm{\mu}}
\,,\nn \\
&=\tfrac{1}{2}\Big(R^{ab}(\omega)+\eta\,s^{a}\wedge s^{b}-\eta\gamma^{-2}\,u^a\wedge u^b\Big)\bm{J}_{ab}
+\Big(\D_\omega s^{a}-\eta\gamma^{-2}u^{a}\wedge\mu\Big)\bm{B}_{a} 
\nn\\
&~~~~~~~~~~~+\Big(\D_{\omega} u^{a}-\eta\,s^{a}\wedge\mu\Big)\bm{T}_{a}
+\Big(\D_{\omega}\mu-\eta_{ab}\,s^{a}\wedge u^{b}\Big)\bm{D}\,,
\end{align}
where
\begin{equation}
\bm{R}(\bm{\omega})=\tfrac{1}{2}R^{ab}(\omega)\bm{J}_{ab}\,,\qquad R^{ab}(\omega)=\d\omega^{ab}+`\omega^a_c`\wedge\omega^{cb}\,,
\end{equation}
is the Lorentz curvature with respect to the Lorentz spin-connection $\omega^{\alpha\beta}$ and
\begin{align}
\D_\omega u^{a}&=\d u^{a}+`\omega^a_b`\wedge u^{b}\,,\qquad
\D_\omega s^{a}=\d s^{a}+`\omega^a_b`\wedge s^{b}\,,\qquad
\D_\omega\mu=\d\mu\,.
\end{align}
\end{itemize}

\subsection{Gauge transformations}

The gauge connection transforms infinitesimally as $\delta\bm{A}=\D_{\bm{A}}\bm{\Lambda}$.
In the following we write out these transformations in components of the different bases.

\subsubsection{4-covariant basis}\label{app:gt4c}

In the 4-covariant basis we denote the 0-form gauge parameter as
\begin{equation}
\bm{\Lambda}=\tfrac{1}{2}\theta^{ab}\bm{J}_{ab}+\beta^{a}\bm{B}_{a}+\tau^{a}\bm{T}_{a}+\lambda\bm{D}\,.
\end{equation}
The transformations of the components of the gauge connection then read
\begin{subequations}\label{eq:gauget4covb} 
\begin{align}
\label{eq:gauget4covb1} 
\tfrac{1}{2}\delta\omega^{ab}&=\tfrac{1}{2}\D_\omega\theta^{ab}+\eta\,s^{[a}\wedge\beta^{b]}
-\gamma^{-2}u^{[a}\wedge \tau^{b]}\,,\\
\label{eq:gauget4covb2} 
\delta s^{a}&=\D_{\omega}\beta^{a}-`\theta^{a}_{b}`\wedge s^{b}
+\eta\gamma^{-2}\Big(\tau^{a}\wedge\mu-u^{a}\wedge\lambda\Big)\,,\\
\label{eq:gauget4covb3} \delta u^{a}
&=\D_{\omega}\tau^{a}-`\theta^{a}_{b}`\wedge u^{b}
+\eta\,\Big(\beta^{a}\wedge\mu-s^{a}\wedge\lambda\Big)\,,\\
\label{eq:gauget4covb4} 
\delta \mu&=\D_{\omega}\lambda-\eta_{ab}\Big(s^{(a}\wedge \tau^{b)}-\beta^{(a}\wedge u^{b)}\Big)\,.
\end{align}
\end{subequations}
It is easy to verify that we recover the gauge symmetries $\ISO(3,2)$ 
or $\ISO(4,1)$ for $\eta=\pm1$ respectively, 
when we take the contraction limit $\gamma\rightarrow\infty$ in \eqref{eq:gauget4covb},
\begin{subequations}\label{eq:gtpc5d}
\begin{align}
\label{eq:gauget4covb1p} \tfrac{1}{2}\delta\omega^{ab}&=\tfrac{1}{2}\D_\omega\theta^{ab}+\eta\,s^{[a}\wedge\beta^{b]}\,,\\
\label{eq:gauget4covb2p} 
\delta s^{a}&=\D_{\omega}\beta^{a}-`\theta^{a}_{b}`\wedge s^{b}\,,\\
\label{eq:gauget4covb3p} 
\delta u^{a}&=\D_{\omega}\tau^{a}-`\theta^{a}_{b}`\wedge u^{b}
+\eta\,\Big(\beta^{a}\wedge\mu-s^{a}\wedge\lambda\Big)\,,\\
\label{eq:gauget4covb4p} 
\delta \mu&=\D_{\omega}\lambda-\eta_{ab}\Big(s^{(a}\wedge \tau^{b)}-\beta^{(a}\wedge u^{b)}\Big)\,.
\end{align}
\end{subequations}

\subsubsection{4-covariant canonical basis}\label{app:gt4cc}

In the 4-covariant canonical basis, we express the gauge parameters as follows,
\begin{equation}
\bm\Lambda=\bm{\theta}+\bm{\rho}+\bm{b}+\bm{\lambda}=\tfrac{1}{2}\theta^{ab}\bm{J}_{ab}+\rho^{a}\bm{P}_{a}+b^{a}\bm{K}_{a}+\lambda\bm{D}.
\end{equation}
Then the gauge transformations for the components in this basis are given by
\begin{subequations}\label{eq:gauget4covcanb}
\begin{align}
\nn \label{eq:gauget4covcanb1}\tfrac{1}{2}\delta\omega^{ab}
&=\tfrac{1}{2}\D_\omega\theta^{ab}+\eta\p{a^2-b^2\gamma^{-2}}e^{[a}\wedge\rho^{b]}\\
 &-\eta\p{ac-bd\gamma^{-2}}\Big(\rho^{[a}\wedge h^{b]}-e^{[a}\wedge b^{b]}\Big)
 +\eta\p{c^2-d^2\gamma^{-2}}h^{[a}\wedge b^{b]}\,,\\
\nn \label{eq:gauget4covcanb2} \delta e^{a}&=\D_{\omega}\rho^{a}-`\theta^{a}_{b}`\wedge e^{b}\\
&-\tfrac{\eta}{\det M}\Big[\Big(ac-bd\gamma^{-2}\Big)\Big(\rho^{a}\wedge\mu-e^{a}\wedge\lambda\Big)
+\Big(c^2-d^2\gamma^{-2}\Big)\Big(b^{a}\wedge\mu-h^{a}\wedge\lambda\Big)\Big]\,,\\
\nn \label{eq:gauget4covcanb3}  \delta h^{a}&=\D_{\omega}b^{a}-`\theta^{a}_{b}`\wedge h^{b}\\
&+\tfrac{\eta}{\det M}\Big[\Big(a^2-b^2\gamma^{-2}\Big)\Big(\rho^{a}\wedge\mu-e^{a}\wedge\lambda\Big)
+\Big(ac-bd\gamma^{-2}\Big)\Big(b^{a}\wedge\mu-h^{a}\wedge\lambda\Big)\Big]\,,\\
\label{eq:gauget4covcanb4} 
\delta \mu&=\D_{\omega}\lambda-\det M\,\eta_{ab}\Big(e^{(a}\wedge b^{b)}-\rho^{(a}\wedge h^{b)}\Big)\,.
\end{align}
\end{subequations}

\subsubsection{Orthogonal basis}\label{app:gt4dc}

Finally, we note that the gauge transformations in the orthogonal basis are given by
\begin{subequations}\label{eq:gtdcb}
\begin{align}
\tfrac{1}{2}\delta\omega^{ab}&=\tfrac{1}{2}\D_\omega\theta^{ab}+\eta A\, e^{[a}\wedge\rho^{b]}+\eta\,C\,h^{[a}\wedge b^{b]}\,,\\
\delta e^{a}&=\D_{\omega}\rho^{a}-`\theta^{a}_{b}`\wedge e^{b}-\tfrac{\eta C}{\det M}\Big(b^{a}\wedge\mu-h^{a}\wedge\lambda\Big)\,,\\
\delta h^{a}&=\D_{\omega}b^{a}-`\theta^{a}_{b}`\wedge h^{b}+\tfrac{\eta A}{\det M}\Big(\rho^{a}\wedge\mu-e^{a}\wedge\lambda\Big)\,,\\
\delta \mu&=\D_{\omega}\lambda-\det M\,\eta_{ab}\Big(e^{(a}\wedge b^{b)}-\rho^{(a}\wedge h^{b)}\Big)\,,
\end{align}
\end{subequations}

\section{Chern-Simons action in components}

\paragraph{4-covariant canonical basis.}
In the components of the 4-covariant canonical basis of equation (\ref{connsu}) 
the 5-dimensional Chern-Simons action reads
\begin{align}\label{eq:thebigaction} 
\nn &S_{\text{CS}_5}[\omega,e,h,\mu]\\
\nn &=\tfrac{1}{4}\int_{M_5}\epsilon_{abcd}\,R^{ab}(\omega)\wedge R^{cd}(\omega)\wedge\mu\\
\nn &+\tfrac{\eta}{2}\int_{M_5}\epsilon_{abcd}\Big(A\,R^{ab}(\omega)\wedge e^{c}\wedge e^{d}+2B\,R^{ab}(\omega)\wedge e^{c}\wedge h^{d}+C\,R^{ab}(\omega)\wedge h^{c}\wedge h^{d}\Big)\wedge\mu\\
\nn &+\tfrac{1}{4}\int_{M_5}\epsilon_{abcd}
\Big(A^2\,e^{a}\wedge e^{b}\wedge e^{c}\wedge e^{d}+4AB\,e^{a}\wedge e^{b}\wedge e^{c}\wedge h^{d}
+2(AC+2B^2)\,e^{a}\wedge e^{b}\wedge h^{c}\wedge h^{d}
\\
\nn &\hspace{160pt}
+4BC\,e^{a}\wedge h^{b}\wedge h^{c}\wedge h^{d}
+C^2\,h^{a}\wedge h^{b}\wedge h^{c}\wedge h^{d}\Big)\wedge\mu\\
\nn &-\int_{M_5}\epsilon_{abcd}\,R^{ab}(\omega)\wedge
\Big(ab\,T^{c}(\omega,e)\wedge e^{d}+ad\,T^{c}(\omega,e)\wedge h^{d}
+bc\,T^{c}(\omega,h)\wedge e^{d}+cd\,T^{c}(\omega,h)\wedge h^{d}\Big)\\
\nn &-\eta\int_{M_5}\epsilon_{abcd}\,\Big(a\,T^{a}(\omega,e)+c\,T^{a}(\omega,h)\Big)\wedge
\Big(b\p{A+\tfrac{2}{3}b^2\gamma^{-2}}e^{b}\wedge e^{c}\wedge e^{d}\\
\nn & \hspace{40pt}
+\p{d\p{A+\tfrac{2}{3}b^2\gamma^{-2}}
+2b\p{B+\tfrac{2}{3}bd\gamma^{-2}}}e^{b}\wedge e^{c}\wedge h^{d}\\
&\hspace{40pt}
+\p{b\p{C+\tfrac{2}{3}d^2\gamma^{-2}}
+2d\p{B+\tfrac{2}{3}bd\gamma^{-2}}}e^{b}\wedge h^{c}\wedge h^{d}
+d\p{C+\tfrac{2}{3}d^2\gamma^{-2}}h^{b}\wedge h^{c}\wedge h^{d}\Big)\,,
\end{align}
where we have defined the parameter combinations
\begin{equation}
A=a^2-b^2\gamma^{-2}\,,\qquad B=ac-bd\gamma^{-2}\,,\qquad C=c^2-d^2\gamma^{-2}\,.
\end{equation}

\paragraph{Conformal basis.} The torsion terms in the conformal basis in equation (\ref{eq:csgamma0case1}) 
are given by
\begin{align}\label{torsionterms}
\nn S_{\text{torsion}}[\omega,e,h]
&=\mp\int_{M_5}\epsilon_{abcd}\,R^{ab}(\omega)\wedge
\Big(a^2\,T^{c}(\omega,e)\wedge e^{d}-ac\,T^{c}(\omega,e)\wedge h^{d}\\
\nn &\qquad\qquad\qquad
+ac\,T^{c}(\omega,h)\wedge e^{d}-c^2\,T^{c}(\omega,h)\wedge h^{d}\Big)\\
\nn &
\mp\frac{2\eta}{3}\int_{M_5}\epsilon_{abcd}\,\Big(a\,T^{a}(\omega,e)+c\,T^{a}(\omega,h)\Big)\wedge
\Big(a^3\,e^{b}\wedge e^{c}\wedge e^{d}\\
&\qquad\qquad\qquad
+3a^2c\,e^{b}\wedge e^{c}\wedge h^{d}-3ac^2\,e^{b}\wedge h^{c}\wedge h^{d}-c^3\,h^{b}\wedge h^{c}\wedge h^{d}\Big)\,,
\end{align}
for the algebraic solutions of $A=C=0$ given by $b=\pm a\gamma$ and $d=\mp c\gamma$ respectively.  

\paragraph{Orthogonal basis.} In the orthogonal basis, 
i.e.~for $b=\pm ia\gamma$ and $d=\mp ic\gamma$, the torsion terms of equation \eqref{eq:csgamma0case3} are
\begin{align}\label{torsionterms2}
\nn S'_{\text{torsion}}[\omega,e,h]
&=\mp i\int_{M_5}\epsilon_{abcd}\,R^{ab}(\omega)\wedge
\Big(a^2\,T^{c}(\omega,e)\wedge e^{d}-ac\,T^{c}(\omega,e)\wedge h^{d}\\
\nn &\qquad\qquad\qquad
+ac\,T^{c}(\omega,h)\wedge e^{d}-c^2\,T^{c}(\omega,h)\wedge h^{d}\Big)\\
& \pm i\frac{4\eta}{3}\int_{M_5}\epsilon_{abcd}\,\Big(a\,T^{a}(\omega,e)+c\,T^{a}(\omega,h)\Big)\wedge
\Big(a^3\,e^{b}\wedge e^{c}\wedge e^{d}-c^3\,h^{b}\wedge h^{c}\wedge h^{d}\Big)\,.
\end{align}

\bibliography{bibtexfile.bib} 
\bibliographystyle{ieeetr}

\end{document}